\documentclass[preprint,12pt]{elsarticle}

\usepackage{geometry}
\usepackage{graphics,graphicx}
\usepackage[usenames,dvipsnames]{xcolor}
\usepackage{amsmath}
\usepackage{latexsym}
\usepackage{amsfonts}
\usepackage{mathrsfs}
\usepackage{bigints}
\usepackage{bm}
\usepackage{multirow}
\usepackage{amsthm}
\usepackage{amssymb}
\usepackage{caption}
\usepackage{subcaption}
\usepackage{epstopdf}
\usepackage{float}
\usepackage[colorlinks,citecolor=blue,urlcolor=blue,bookmarks=false,hypertexnames=true]{hyperref}
\usepackage[title]{appendix}
\usepackage{tikz}
\usetikzlibrary{shapes.geometric, arrows}
\tikzstyle{startstop} = [rectangle, rounded corners, minimum width=3cm, minimum height=1cm,text centered, draw=black]
\tikzstyle{arrow} = [thick,->,>=stealth]
\graphicspath{ {Fig/} }

\usepackage[labelformat=simple]{subcaption}

\newtheorem{theorem}{Theorem}[section]

\usepackage{tikz}

\usetikzlibrary{shapes.geometric, arrows, shapes}
\tikzstyle{startstop} = [rectangle, rounded corners, minimum width=3cm, minimum height=1cm,text centered, draw=black]
\tikzstyle{arrow} = [thick,->,>=stealth]

\newcommand\solidrule[1][1cm]{\rule[0.6mm]{6mm}{2pt}}

\definecolor{fadblu}{rgb}{0, 0.4470, 0.7410}
\definecolor{darkmar}{rgb}{0.6350, 0.0780, 0.1840}
\definecolor{mag}{rgb}{1.00,0.07,0.65}

\definecolor{fadblu1}{rgb}{0.2, 0.40, 0.76}
\definecolor{mag1}{rgb}{0.8,0.07,0.7}
\definecolor{gray1}{rgb}{0.5,0.5,0.5}

\definecolor{GRN}{rgb}{0.4660, 0.6740, 0.1880}
\definecolor{MRN}{rgb}{0.6350, 0.0780, 0.1840}

\biboptions{sort&compress}

\makeatletter

\newcommand{\mathleft}{\@fleqntrue\@mathmargin0pt}
\newcommand{\mathcenter}{\@fleqnfalse}
\makeatother

\usetikzlibrary{positioning, fit, calc}   
\tikzset{block/.style={draw, thick, text width=2cm ,minimum height=1.3cm, align=center},   
line/.style={-latex}     
}

\begin{document}

\begin{frontmatter}


\title{Analysis of the impact of fear in the presence of additional food and prey refuge with nonlocal predator-prey models}

\author[inst1]{Sangeeta Saha\corref{cor1}}
\ead{sangeetasaha629@gmail.com}
\author[inst1]{Swadesh Pal}
\ead{spal@wlu.ca}
\author[inst1,inst2]{Roderick Melnik}
\ead{rmelnik@wlu.ca}

\cortext[cor1]{Corresponding author}
\address[inst1]{MS2 Discovery Interdisciplinary Research Institute, Wilfrid Laurier University, Waterloo, Canada}
\address[inst2]{BCAM - Basque Center for Applied Mathematics, E-48009, Bilbao, Spain}

\begin{abstract}
There are many positive and negative factors present in the predator-prey interaction which affect the net growth of the species. Fear of predation is one such factor that creates psychological stress in a prey species, which causes a negative impact on their overall growth. This work considers a predator-prey model where the prey species faces a reduction in their growth out of fear, and the predator has an alternative food source that helps the prey to hide in a safer place. As an extension into the nonlocal spatio-temporal model, a nonlocal term is considered in the prey growth to incorporate a fear-effect range around their spatial location. Linear stability analysis helps to analyze the temporal model and produces a wide range of interesting results, including the presence of a certain amount of fear or even prey refuge, which helps in population coexistence. Furthermore, the numerical simulations of the local and nonlocal spatio-temporal models show different types of spatial-temporal patterns, such as Turing and non-Turing patterns. Nevertheless, an increase in fear level reduces the range of the Turing domain for the local model, whereas the opposite happens when the range of nonlocal interaction is increased.
\end{abstract}


\end{frontmatter}


\section{Introduction} \label{sec:1}

The essential component in ecology is the interaction between predators and their prey, as this upholds the biomass flow from one trophic level to others and maintains the population size. Researchers have developed several mathematical models to explore the dynamic nature, such as predators searching for prey for survival, prey adopting different strategies to avoid higher predation, etc. This paper considers a predator-prey model where the growth of a prey species is affected by its predator along with their spatial local and nonlocal interactions. \par

In an ecosystem, predators impact prey both directly, through consumption and indirectly, by inducing fear that alters prey behaviour \cite{lima1990behavioral}. This fear creates continuous psychological stress, sometimes leading to greater population declines than direct predation alone \cite{zanette2011perceived}. Prey species exhibit various anti-predator responses, including habitat shifts, increased vigilance, and altered foraging and reproductive strategies \cite{cresswell2011predation, peacor2013costs, preisser2008many}. They may move to safer areas with lower predation risk but sufficient food availability \cite{ripple2004wolves, wirsing2007living}. While these adaptations enhance short-term survival, they often reduce reproductive success over time \cite{cresswell2011predation}.\par

Recently, researchers have focused on pursuing their studies on how indirect methods become more effective than direct killing in reducing prey populations \cite{suraci2016fear, wang2016modelling, zanette2011perceived}. It has been observed that the reproduction of song sparrows was affected by the fear effect of their predators (raccoon, owl, hawk, etc.) during the entire breeding season, even if the direct predation is excluded \cite{zanette2011perceived}. There were fewer eggs, hatchlings, and even fledglings in the next generation, and the reproduction rate decreased by about 40\%. An experiment on free-living wild songbird populations shows that in the presence of predator playback, songbird parents generated 53\% fewer recruits to the adult breeding population \cite{allen2022fear}. On the other hand, parental behaviours of pumpkinseed (\textit{Lepomis gibbosus}), including in-nest rotations and egg and nest maintenance, are altered because of their avian predator, the Osprey (\textit{Pandion haliaetus}) \cite{gallagher2016avian}. Moreover, mesocarnivores (raccoons) reduce their foraging activities by 66\% due to the fear of large carnivores (cougar, wolf, black bear) \cite{suraci2016fear}. In the aquatic environment, it is observed that the normal growth rate of bluegill can rise by almost 27\% when predators, including largemouth bass, trout, turtles, etc., are absent \cite{wootton2012ecology}. Furthermore, there are several predator-prey interactions where the reproduction of the scared prey decreases due to fear of predation risk, e.g., bluebirds- avian predators \cite{hua2014increased}, elk-wolves \cite{creel2007predation, wirsing2011comparison}, snowshoe hares-dogs \cite{sheriff2009sensitive}, dugongs-sharks \cite{wirsing2011comparison}, mule deer-mountain lions \cite{pierce2004habitat}, etc. Research has shown that the anxiousness of a predator alone may alter a prey's behaviour. Wang et al. first included fear of the predator on a prey species in their model by lowering the prey's birth rate \cite{wang2016modelling}. They noticed that the system's oscillation may be stabilized by the fear effect. Several research works have already been conducted to explore how this fear of predation makes an impact on the dynamic behaviour of predator-prey interactions \cite{panday2018stability, zhang2022impact, xie2022influence}, tri-tropic food chain models \cite{saha2021analysis, thirthar2022fear}, ecological models with prey refuge \cite{zhang2019impact}, stage-structured models \cite{wang2017modeling}, spatio-temporal models \cite{sarkar2023spatiotemporal, pal2024cross}, etc. \par

The functional response represents the predator's behavioural traits in predator-prey interactions, significantly influencing dynamic behaviour. Key factors affecting functional responses include prey biomass, predator efficiency in locating and handling prey, and competition. Common prey-dependent responses include Lotka–Volterra and Holling-type models \cite{holling1959components}. However, in many cases, functional responses should be predator-dependent, particularly when predators compete for food. According to a wide range of literature, predator dependency in the functional response appears to be common in both natural systems and laboratory settings \cite{arditi1990underestimation, dolman1995intensity}. Prey alter their behaviour due to predation risk, while predators interfere with each other, leading to competitive effects influenced by prey availability, environmental conditions, or territorial disputes \cite{skalski2001functional}. To mediate between theoretical and experimental viewpoints, Beddington \cite{beddington1975mutual} and DeAngelis et al. \cite{deangelis1975model} independently suggested a functional response considering the mutual interference amongst predators. This model assumes that two or more predator populations spend some time interfering with one another's actions except when searching for and processing the prey species. In addition, the influence of predator interference on the feeding rate decreases with prey quantity, indicating that predators' collective behaviour becomes less significant. \par

Predator fear can restrict prey growth, prompting them to look for alternative food sources. Providing additional food to predators can mitigate this effect in two ways. First, extra food can reduce predation rates by diverting predator attention, allowing prey to find refuge, and lowering extinction risks. For example, in an experiment, it was observed that when a sufficient amount of additional food was given to hen harriers, the predation rate decreased from 3.7 chicks to 0.5 chicks per 100 hours \cite{moorland1999diversionary}. Second, increased food availability may enhance predator vigilance or reproduction rate, which intensifies predation and reduce prey populations. Crawley emphasized that food quality and quantity are critical factors \cite{crawley2009plant}. Several studies state that good quality extra food sources result in high predation rates, whereas low quality may conserve the target prey \cite{huxel2002effects, srinivasu2011role, akhtar2024dynamical}. Proper food supplementation to predators helps maintain ecological balance and prevents excessive prey depletion.\par 

The assumption of homogeneous distribution in temporal models of interacting populations portrays the random movement of individuals. Spatio-temporal models are necessary to accurately capture these dynamics. Since Alan Turing's pioneering work on chemical morphogenesis \cite{turing1990chemical}, reaction-diffusion equations have been widely studied for their role in generating spatio-temporal patterns in predator-prey interactions. The reaction kinetics significantly influence the formation of both stationary and non-stationary spatial patterns. These systems also model ecological invasions and disease spread \cite{sivasamy2019spatial, cantrell2009modeling}. Researchers have identified key mechanisms driving pattern formation, including Hopf bifurcation \cite{fussmann2000crossing} and diffusion-driven instabilities \cite{PhysRevE.75.051913, sarkar2023spatiotemporal}. Non-stationary patterns, such as chaotic patterns and travelling waves, prevent population patches from stabilizing. Stationary patterns typically emerge within the Turing or Turing-Hopf domain, while non-stationary patterns appear in specific parameter ranges beyond Turing instability. \par

The spatio-temporal local models are formulated based on the assumption that individuals at a given spatial point interact with one another only and that members of both species can move between spatial points. However, these models cannot account for scenarios where individuals access resources at neighbouring spatial points. To address this, nonlocal interaction terms have been introduced in reaction kinetics \cite{BANERJEE20172, BAYLISS201737, pal2018analysis, pal2024pattern}. In nonlocal predator-prey systems, predators can migrate to exploit distant resources \cite{BANERJEE20172}. Notably, such models exhibit spatio-temporal chaos at high consumption rates, unlike their local counterparts, which fail to produce Turing patterns. \par

As mentioned above, fear is a psychological response that alters prey behaviour by reducing feeding time, shifting habitat, and lowering reproduction, yet its ecological impact remains under-explored. Wang et al. \cite{wang2016modelling} first incorporated fear into predator-prey models, showing its stabilizing effects. Subsequent studies have examined fear in various ecological contexts, including multi-trophic systems, prey refuge models, and spatio-temporal dynamics \cite{panday2018stability, zhang2022impact, xie2022influence, saha2021analysis, thirthar2022fear, zhang2019impact, wang2017modeling, sarkar2023spatiotemporal}. On the contrary, providing additional food to predators helps maintain ecological balance by preventing prey overexploitation. Research has analyzed its effects in fractional-order systems highlighting memory effects \cite{debnath2023memory}, stochastic models \cite{das2018modeling, das2020prey}, and density-dependent functional responses \cite{pal2019role}. Additionally, studies have explored predator-prey dynamics with prey refuge and additional food provision \cite{thirthar2022fear, ghosh2017prey} in shaping ecosystem stability \cite{thirthar2022fear, ghosh2017prey}.

Our main intention in this work is to explore the impact of psychological effects affecting predator-prey interactions in terms of fear of predation through a local and nonlocal approach. Here, a predator-prey relationship is presented, in which the fear of the predators affects the prey's reproduction, and the predators interfere in their activities during prey consumption. In this case, there is an indirect psychology that works among predators to get more food. In addition, the prey species has a tendency to move towards a predator-free zone in the absence of fear. 
While existing literature addresses predator-prey interactions incorporating factors such as fear effects, prey refuge, and additional food resources separately, to the best of our knowledge, no prior studies have concurrently considered all of these elements within a unified framework. Investigating this combined approach would provide a more comprehensive understanding of the dynamic nature of such systems. Furthermore, the inclusion of both local and nonlocal interactions mediated by the fear term represents a distinctive and novel contribution to the current manuscript.
We have revealed several results, such as the fear of predation affecting the coexistence of the species in an environment, the influence of other psychological factors, including intra-specific competition and prey refuge on the population, etc. Nevertheless, we have shown that the nonlocal fear term positively impacts species migration. 


\section{Model Formulation} \label{sec:2}

The environment mainly consists of food webs and corresponding food chains, which are nothing but predator-prey interactions. Many factors present in our surroundings contribute positively or negatively to the growth of a species. In this work, we are dealing with a predator-prey interaction where some psychological aspects for both species have been considered. Wang et al., in 2016, have proposed a predator-prey model where the growth rate of prey is reduced because of the fear of predation \cite{wang2016modelling}. The model they have proposed is given as:
\begin{equation}\label{eq:eq1}
\begin{aligned}
\frac{du}{dt}&=ru\overline{f}_{1}(\omega,v)-du-pu^{2}-\overline{f}_{2}(u)v,\ \ u(0)>0, \\
\frac{dv}{dt}&=v(-\mu_{1}+c\overline{f}_{2}(u)),\ \ v(0)>0,
\end{aligned}  
\end{equation} 
where $\omega$ defines the level of fear that drives the anti-predator behaviour of prey, and the function $\overline{f}_{1}(\omega,v)=1/(1+\omega v)$ satisfies the following conditions:
\begin{align*}
(i)\ \overline{f}_{1}(\omega,0)=1, \quad (ii)\ \overline{f}_{1}(0,v)&=1 \quad (iii)\ \lim_{\omega\rightarrow\infty}\overline{f}_{1}(\omega,v)=0, \quad (iv)\ \lim_{v\rightarrow\infty}\overline{f}_{1}(\omega,v)=0, \\ 
(v)\ \frac{\partial\overline{f}_{1}}{\partial\omega}&<0, \quad \mbox{and}\quad (vi)\ \frac{\partial\overline{f}_{1}}{\partial v}<0.
\end{align*}

The prey and predator biomass are denoted by $u$ and $v$, respectively. The growth rate of prey species is $r$, and the parameter $d$ is their natural death rate. The prey is involved in intra-specific competition at rate $p$. Lastly, $\overline{f}_{2}(u)$ denotes the prey-dependent functional response, and $\mu_{1}$ is the natural mortality rate of predator. Though they have considered only prey-dependent functional response, there is literature stating that the predators affect each other's activities when they become large in number compared to the available prey biomass \cite{skalski2001functional}. In this work, we have chosen a predator-dependent functional response, named the Beddington-DeAngelis response, where it is considered that the predator species spend some time encountering each other except searching for and processing the prey. The Beddington–DeAngelis functional response is considered for modelling predation primarily due to its generality. Notably, the Holling type II response emerges as a special case when there is no mutual interference among predators. Predator-prey models incorporating the Beddington-DeAngelis functional response exhibit more intricate dynamical behaviour compared to those based on the Holling type II response. This complexity arises because, during the predation process, the time lost due to encounters between predators serves as a key factor in deriving the Beddington–DeAngelis functional response.
Let us consider $t_{u}$ and $t_{v}$ as the handling times of the predator per prey and interaction time between two or more predators. Then, the maximum predation rate will be $a=1/t_{u}$. Moreover, the parameters $e_{u}$ and $e_{v}$ are the constants that signify the rate of the predator's movement to detect a prey and another predator, respectively \cite{cantrell2004spatial}. Then the parameter $b=1/(e_{u}t_{u})$ denotes the normalization coefficient relating the prey and predator biomass to their interacting environment \cite{deangelis1975model}. Here, $c$ is a quantity from $(0,1)$ that represents the positive contribution of the food consumed by the predator, converted into predator biomass, which makes `$ca$' the maximum growth of predator species. And lastly, as we have considered the predator's interference at the time of consumption, the parameter $l$ looks like $l=(e_{v}t_{v})/(e_{u}t_{u})$. So, the predator-prey interaction stated in (\ref{eq:eq1}) with the Beddington-DeAngelis functional response takes the form \cite{thirthar2022fear, zhang2022impact}: 
\begin{equation}\label{eq:eq2}
\begin{aligned}
\frac{du}{dt}&=ru\overline{f}_{1}(\omega,v)-du-pu^{2}-\frac{auv}{b+u+lv},\ \ u(0)>0, \\
\frac{dv}{dt}&=v\left(-\mu_{1}+\frac{cau}{b+u+lv}\right),\ \ v(0)>0.
\end{aligned}  
\end{equation} 

Now, a predator species usually does not depend only on one particular prey, so let us assume an additional/alternative food source of biomass $A$ is uniformly distributed to the habitat where the predator and prey interact. It is considered that the amount of additional food is proportional to the number of encounters the predator is having with additional food. Let the parameter $\alpha=t_{A}/t_{u}$ signify the quality of additional food compared to the prey species if $t_{A}$ is chosen to be the time taken by the predator to handle per unit of additional food. Also, if $e_{A}$ is considered as the constant representing the rate of movement of the predator at the time of searching to detect the alternative food source, then $\eta=e_{A}/e_{u}$ represents the effectual ability of the predator species to detect the additional food. It means $\eta A$ is the quantity of additional food the predator can notice relative to the prey species. Hence, in this case, the model becomes
\begin{equation}\label{eq:eq3}
\begin{aligned}
\frac{du}{dt}&=ru\overline{f}_{1}(\omega,v)-du-pu^{2}-\frac{auv}{b+\alpha\eta A+u+lv},\ \ u(0)>0, \\
\frac{dv}{dt}&=v\left(-\mu_{1}+\frac{ca(u+\eta A)}{b+\alpha\eta A+u+lv}\right),\ \ v(0)>0.
\end{aligned}  
\end{equation} 

The model (\ref{eq:eq3}) is a predator-prey interaction where the growth rate of prey species is affected by fear of predators, additional food is provided to predators, and the predators follow Beddington-DeAngelis functional response towards available food. 

When the predator spends time searching and handling alternative food sources, the targeted prey can successfully hide in places that are not accessible to the predator, creating a prey refuge. Here, it is considered that $m$ is the fraction of prey hiding, which leaves $(1-m)$ portion of prey provided to the predator. Now, if many predators are exposed to prey species that are inadequate in the environment, they become involved in intra-specific competition. Gaining more food becomes the priority in such a scenario. We have considered parameter $\mu_{2}$ as the intra-specific competition rate of predator species. The parameter $l$ is associated with predator interference during consumption, while $\mu_{2}$ signifies intra-specific competition among predators that arises from factors other than direct consumption \cite{hsu2001rich, bazykin1998nonlinear}. 
Such competition may stem from habitat limitations, territorial conflicts, mating competition, resource scarcity, or dominance hierarchies among predators. Hence, summing up all the assumptions, we finally propose the temporal model as follows:
\begin{equation}\label{eq:det1}
\begin{aligned}
\frac{du}{dt}&=\frac{ru}{1+\omega v}-du-pu^{2}-\frac{a(1-m)uv}{b+\alpha\eta A+(1-m)u+lv}\equiv \frac{ru}{1+\omega v}-uf_{1}(u,v),\\
\frac{dv}{dt}&=\frac{ca\{(1-m)u+\eta A\}v}{b+\alpha\eta A+(1-m)u+lv}-\mu_{1}v-\mu_{2}v^{2}\equiv vf_{2}(u,v),
\end{aligned}  
\end{equation} 
with non-negative initial conditions. The system parameters are chosen to be positive along with $r>d$. 

\subsection{Local spatio-temporal model} \label{sec:3}

The distributions of populations are generally heterogeneous and depend on time and the spatial positions in the habitat. So, it is natural and more precise to study the corresponding PDE problem. In this work, we have considered the spatio-temporal model with reaction kinetics of system (\ref{eq:det1}) in a bounded domain $\Omega\subset \mathbb{R}^{n}$ for $n=1,2$ with closed boundary $\partial\Omega$ and $\overline{\Omega}=\Omega\cup \partial\Omega$ as:
\begin{equation} \label{eq:diff1}
\begin{aligned}
\frac{\partial u}{\partial t}&=d_{1}\Delta u+\frac{ru}{1+\omega v}-du-pu^{2}-\frac{a(1-m)uv}{b+\alpha\eta A+(1-m)u+lv},\\
\frac{\partial v}{\partial t}&=d_{2}\Delta v+\frac{ca\{(1-m)u+\eta A\}v}{b+\alpha\eta A+(1-m)u+lv}-\mu_{1}v-\mu_{2}v^{2},
\end{aligned}
\end{equation}
subject to non-negative initial conditions. To make the model simple, we have chosen periodic boundary conditions as there is not much qualitative change in the dynamics of the nonlocal model by changing the boundary conditions periodic to no-flux \cite{pal2020effects}. Here, $\Delta\equiv\partial^{2}/\partial x^{2}$, and the parameters $d_{1}$ and $d_{2}$ represent the diffusion coefficients corresponding to prey and predator species. In addition, instead of one-dimensional diffusion, if the species move in a two-dimensional domain, then $\Delta\equiv\partial^{2}/\partial x^{2}+\partial^{2}/\partial y^{2}$. In this case, the closed bounded domain becomes $\Omega\subset \mathbb{R}^{2}$ with a closed boundary $\partial\Omega$ and $\overline{\Omega}=\Omega\cup \partial\Omega$. 

\subsection{Nonlocal interaction} \label{sec:4}

In one spatial dimension, the model (\ref{eq:diff1}) assumes that the predator located at the space point $x$ impacts the growth of prey at the same point. However, in reality, the fear of predators at a spatial location $x$ depends not only on the local appearance of the predator but also on the predator density in other nearby points, i.e., a prey located at space point $x$ can be scared by those predators who are located in some areas around this spatial point, and it can be obtained by convolution term
$$(\phi_{\delta}*v)(x,t)=\int_{-\infty}^{\infty}\phi_{\delta}(x-y)v(y,t)dy.$$

Here, the kernel function $\phi_{\delta}(y)$ describes the impact of fear on prey at the space point $x$ by the predator at the space point $y$. Hence, the kernel $\phi_{\delta}$ is a function dependent on the position $x$. The first subscript $\delta$ is the range of nonlocal interaction. We assume the kernel function $\phi_{\delta}$ to be non-negative with compact support. Also, to preserve the same homogeneous steady-state solutions for both the local and nonlocal models, we assume that $\int_{-\infty}^{\infty}\phi_{\delta}(y)dy=1$. Several types of kernel functions have been studied while dealing with nonlocal models, such as top-hat, parabolic, triangular, and Gaussian \cite{pal2020effects, BAYLISS201737, pal2024nonlocal}. However, we have considered the top-hat kernel function in this work. The impact of fear on prey is limited by the number of predators they are facing or encountering. We can apply this limitation to each space point independently. This means that predators located at space point $y$ make an impact on prey at space point $x_{1}$ independently of its concentration at another point $x_{2}$. Considering the work of Furter and Grinfeld \cite{furter1989local} and using the aforementioned assumption, we obtain the rate of impact of fear on prey at the space point $x$ as
$$\frac{ru}{1+\omega(\phi_{\delta}*v)}=\frac{ru(x,t)}{1+\omega\int_{-\infty}^{\infty}\phi(x-y)v(y,t)dy}.$$
Implementing the nonlocal fear term of prey species as well as the random motion of the population, we get the integro-differential equation model as
\begin{equation} \label{eq:loc1}
\begin{aligned}
\frac{\partial u}{\partial t}&=d_{1}\frac{\partial^{2}u}{\partial x^{2}}+\frac{ru}{1+\omega(\phi_{\delta}*v)}-du-pu^{2}-\frac{a(1-m)uv}{b+\alpha\eta A+(1-m)u+lv}, \\
\frac{\partial v}{\partial t}&=d_{2}\frac{\partial^{2}v}{\partial x^{2}}+\frac{ca\{(1-m)u+\eta A\}v}{b+\alpha\eta A+(1-m)u+lv}-\mu_{1}v-\mu_{2}v^{2},
\end{aligned}
\end{equation}
with non-negative initial conditions and periodic boundary conditions.

\section{Analysis of the Proposed Systems}

It is important to show that the proposed system is biologically well-defined, and for this, we need to check the positivity and uniform boundedness of the system variables. The following theorem states that the model (\ref{eq:det1}) is well-posed, and its proof is given in Appendix \ref{appendixA1}.

\begin{theorem}
Solutions of system (\ref{eq:det1}), starting in $\mathbb{R}_{+}^{2}$, are non-negative for $t>0$ and uniformly bounded provided $r>d$.
\end{theorem}

\subsection{Equilibrium points for the temporal model and their stabilities}

The temporal model (\ref{eq:det1}) has\\
\noindent (i) a trivial equilibrium point $E_{0} = (0,0)$;\\
(ii) two planer equilibrium points $E_{1}= (\overline{u},0)\equiv\left((r-d)/p,0\right)$ and $E_{2} = (0,\widetilde{v})$ where $\widetilde{v}$ is the root of the equation
$$\mu_{2}lv^{2}+(\mu_{1}l+\mu_{2}s)v-(ca\eta A-\mu_{1}s)=0~\mbox{with}~s=b+\alpha\eta A.$$
Here, the feasibility of the equilibrium point $E_{2}$ depends on the condition $ca\eta A>\mu_{1}s$; \\
(iii) the coexisting equilibrium point $E^{*} = (u^{*},v^{*})$, where $$u^{*}=\frac{(s+lv^{*})(\mu_{1}+\mu_{2}v^{*})-ca\eta A}{(1-m)[ca-(\mu_{1}+\mu_{2}v^{*})]}$$ and $v^{*}$ is the root of the following equation 
\begin{align}\label{eq:2.2}
   B_{1}v^{4}+B_{2}v^{3}+B_{3}v^{2}+B_{4}v+B_{5}=0,
\end{align}
with 
$B_{1} =a\omega\mu_{2}[cpl^{2}+\mu_{2}(1-m)^{2}], 
B_{2} =a\mu_{2}(1-m)^{2}(\mu_{2}-2\omega A_{1})+cal[p\mu_{2}(l+\omega A_{4})+\omega\{pA_{3}-d\mu_{2}(1-m)\}], 
B_{3} =ca[l\mu_{2}\{pA_{4}+r(1-m)\}+(l+\omega A_{4})\{pA_{3}-d\mu_{2}(1-m)\}+\omega l\{pA_{2}+d(1-m)A_{1}\}] +a(1-m)^{2}(ca-\mu_{1})\{\omega(ca-\mu_{1})-2\mu_{2}\}, 
B_{4} =ca[A_{4}\{pA_{3}+\mu_{2}(r-d)(1-m)\}+(l+\omega A_{4})\{pA_{2}+d(1-m)A_{1}\}-rl(1-m)A_{1}]
+a(1-m)^{2}A_{1}^{2}, 
B_{5} =caA_{4}\{pA_{2}-(r-d)(1-m)A_{1}\},
A_{1} =ca-\mu_{1}, A_{2}=\mu_{1}s-ca\eta A,  A_{3}=s\mu_{2}+l\mu_{1}$, and $A_{4}=s-\eta A$.

Here, $B_{1}>0$ holds always. So, the number of positive $v^{*}$ will depend on the signs of coefficients $B_{i}$, for $i=2,3,4,5$. For example, equation (\ref{eq:2.2}) will not possess any feasible $v^{*}$ if $B_{i}>0$, for $i=2,\ldots,5$, but can have at most four positive roots if $B_{2},\ B_{4}<0$ and $B_{3},\ B_{5}>0$ hold. Moreover, equation (\ref{eq:2.2}) has at least one positive root if any of the following conditions hold: (i) $B_{i}>0$, $i=2,3,4$ and $B_{5}<0$; (ii) $B_{i}>0$, $i=2,3$ and $B_{j}<0$, $j=4,5$; 
(iii) $B_{2}>0$ and $B_{i}<0$, $i=3,4,5$; (iv) $B_{i}<0$, $i=2,..,5$.

Furthermore, if $ca>(\mu_{1}+\mu_{2}v^{*})$ holds along with positive $v^{*}$, then $u^{*}>0$ when $(s+lv^{*})(\mu_{1}+\mu_{2}v^{*})>ca\eta A$ is satisfied. On the other hand, if $ca<(\mu_{1}+\mu_{2}v^{*})$ holds with $v^{*}>0$, then $(s+lv^{*})(\mu_{1}+\mu_{2}v^{*})<ca\eta A$ has to be fulfilled for the feasibility of $u^{*}$.

Now, we discuss the local stability criterion of the equilibrium points, which can be determined by analyzing the eigenvalues of corresponding Jacobian matrices. Let us denote $s=b+\alpha\eta A$. The Jacobian matrix of system (\ref{eq:det1}) is 
\begin{equation}\ \label{eq:det2}
\textbf{J}=  \left(
\begin{array}{cc}
a_{11} & a_{12}  \\
a_{21} & a_{22}
\end{array}
\right),
\end{equation}
where $a_{11}=\frac{r}{1+\omega v}-\frac{a(1-m)v(s+lv)}{\{s+(1-m)u+lv\}^{2}}-d-2pu,\ a_{12}=-\frac{r\omega u}{(1+\omega v)^{2}}-\frac{a(1-m)u\{s+(1-m)u\}}{\{s+(1-m)u+lv\}^{2}}$, \\
$ a_{21}=\frac{ca(1-m)v\{b+lv+\eta A(\alpha-1)\}}{\{s+(1-m)u+lv\}^{2}}$ and $a_{22}=\frac{ca\{(1-m)u+\eta A\}\{s+(1-m)u\}}{\{s+(1-m)u+lv\}^{2}}-\mu_{1}-2\mu_{2}v$. \\

The following theorem states the local stability property of the boundary equilibrium and interior equilibrium points, and its proof is illustrated in Appendix \ref{appendixA2}.

\begin{theorem} \label{th2.2}
System (\ref{eq:det1}) has several axial and interior equilibrium points, among which 
\begin{enumerate}[(a)]
    \item $E_{0}$ is an unstable equilibrium point.
    \item $E_{1}$ is locally asymptotically stable when $(ca-\mu_{1})(1-m)\overline{u}<\mu_{1}s-ca\eta A$ holds.
    \item $E_{2}$ is locally asymptotically stable when $\omega\{dl+a(1-m)\}\widetilde{v}^{2}+\{ds\omega+a(1-m)-(r-d)l\}\widetilde{v}-(r-d)s>0$ holds.
    \item $E^{*}$ is locally asymptotically stable if the following two conditions hold, i.e., \\
(i) $av^{*}[(1-m)^{2}u^{*}-lc(1-m)u^{*}-lc\eta A]<(pu^{*}+\mu_{2}v^{*})K^{2},$ \\
(ii) $p\mu_{2}K^{4}+[ca\{pl\{(1-m)u^{*}+\eta A\}+r\omega(1-m)\{b+lv^{*}+\eta A(\alpha-1)\}\}-a\mu_{2}(1-m)^{2}v^{*}]K^{2}+ca^{2}(1-m)^{2}[\{s+(1-m)u^{*}\}\{b+lv^{*}+\eta A(\alpha-1)\}+(s-\eta A)lv^{*}]>0$, \\
where $K=[s+(1-m)u^{*}+lv^{*}]>0$.
\end{enumerate}
\end{theorem}



\subsection{Local bifurcations for the temporal model}

The local bifurcations around the equilibrium points are analyzed mainly with the help of Sotomayor's theorem and Hopf's bifurcation theorem \cite{perko2013differential}. For instance, if the stability condition of any of the equilibrium points in the system violates in such a way that the corresponding determinant becomes $0$, giving a simple zero eigenvalue, then there will occur transcritical bifurcation, and we can observe the exchange of stability in that bifurcation threshold. The following theorems will state the conditions where such bifurcation can be observed for $E_{1}$ and $E_{2}$.


\begin{theorem} \label{th-2.6}
Choosing $a$ as the bifurcation parameter, system (\ref{eq:det1}) undergo transcritical bifurcations around 

(i) $E_{1}(\overline{u},0)$ at $\displaystyle a_{[TC_1]}=[\mu_{1}\{s+(1-m)\overline{u}\}]/[c\{\eta A+(1-m)\overline{u}\}]$.

(ii) $E_{2}(0, \widetilde{v})$ at $a=a_{[TC_2]}$, where $a_{[TC_2]}$ is the positive root of the equation $\omega\{dl+a(1-m)\}\widetilde{v}^{2}+\{ds\omega+a(1-m)-(r-d)l\}\widetilde{v}-(r-d)s=0$. 
\end{theorem}


On the other hand, if any of the mentioned inequalities in (\ref{eq:2.4_1}) is violated, then the equilibrium point becomes unstable, and the system performs oscillatory or non-oscillatory behaviour. The system starts to oscillate around $(u^{*}, v^{*})$ if $F_{1}>0$ along with $F_{1}^{2}-4F_{2}<0$ as the eigenvalues will be in the form of the complex conjugate in this case. So, we get the following theorem: 

\begin{theorem} \label{th-2.7}
Suppose $E^{*}$ exists with the feasibility conditions. A simple Hopf bifurcation occurs at unique $\omega=\omega_{[H]}$ around $E^{*}$, where $\omega_{[H]}$ is the positive root of $F_{1}(\omega)=0$ providing $F_{2}(\omega_{[H]})>0$ (stated in equation (\ref{eq:2.4})).
\end{theorem}

The proofs of both Theorems \ref{th-2.6} and \ref{th-2.7} are elaborated in Appendix \ref{appendixA3}.

\subsection{Turing instability analysis for the local and nonlocal models}

We intend to find the condition for Turing instability. If the homogeneous steady state of the temporal model is locally stable to infinitesimal perturbation but becomes unstable in the presence of diffusion, a scenario of Turing instability occurs. For the temporal model, $E^{*}(u^{*},v^{*})$ is locally asymptotically stable when $a_{11}+a_{22}<0$ and $a_{11}a_{22}-a_{12}a_{21}>0$ hold. Here, we apply heterogeneous perturbation around $E^{*}$ to obtain the criterion for instability of the spatio-temporal model. Let us perturb the homogeneous steady state of the local system (\ref{eq:diff1}) around $(u^{*},v^{*})$ by 
\mathcenter
\begin{align*}
    \begin{pmatrix}
        u \\
        v
    \end{pmatrix}=\begin{pmatrix}
        u^{*} \\
        v^{*}
    \end{pmatrix}+\epsilon \begin{pmatrix}
        u_{1} \\
        v_{1}
    \end{pmatrix}e^{\lambda t+ikx}
\end{align*}
with $|\epsilon|\ll 1$. Here, $\lambda$ is the growth rate of perturbation, and $k$ is the wave number. Substituting these values in system (\ref{eq:diff1}), the linearization takes the form:
\begin{equation} \label{eq:3.2}
\textbf{J}_{k}\begin{bmatrix}
u_{1}\\
v{1}
\end{bmatrix}\equiv \begin{bmatrix}
 a_{11}-d_{1}k^{2}-\lambda & a_{12} \\
 a_{21} & a_{22}-d_{2}k^{2}-\lambda 
\end{bmatrix}
\begin{bmatrix}
    u_{1} \\
    v_{1}
\end{bmatrix}
=\begin{bmatrix}
        0 \\
        0
    \end{bmatrix},
\end{equation}
where $a_{11},\ a_{12},\ a_{21}$, and $a_{22}$ are mentioned in the proof of Theorem \ref{th2.2}. We are interested in finding the non-trivial solution of the system (\ref{eq:3.2}), so $\lambda$ must be a root of $\det(\textbf{J}_{k})=0$. Now,
\begin{align*}
    \det(\textbf{J}_{k}) = 0 \Rightarrow \lambda^{2}-B(k^{2})\lambda+C(k^{2}) = 0,
\end{align*}
where $B(k^{2})=\mbox{tr}(\textbf{J}(E^{*}))-(d_{1}+d_{2})k^{2},\ C(k^{2})=\det(\textbf{J}(E^*))-(d_{2}a_{11}+d_{1}a_{22})k^{2}+d_{1}d_{2}k^{4}$. So, $\det(\textbf{J}_{k})=0$ gives 
\begin{align*}
    \lambda_{\pm}(k^{2})=\frac{B(k^{2})\pm\sqrt{(B(k^{2}))^{2}-4C(k^{2})}}{2}.
\end{align*}
Here, $B(k^{2})<0$ for all $k$ when the temporal model is locally asymptotically stable. So, the homogeneous solution will be stable under heterogeneous perturbation when $C(k^{2})>0$ for all $k$. The system is unstable if the inequality is violated for some $k\neq 0$. Now, $\displaystyle k^{2}_{min}=(d_{2}a_{11}+d_{1}a_{22})/2d_{1}d_{2}$ is the minimum value of $k^{2}$ for which $\displaystyle C(k^{2})$ will attain its minimum value, say $\displaystyle C_{min}$, such that
$$\displaystyle C_{min}=(a_{11}a_{22}-a_{12}a_{21})-\frac{(d_{2}a_{11}+d_{1}a_{22})^{2}}{4d_{1}d_{2}}.$$
This $k_{min}$ is the critical wave number for Turing instability. And, the critical diffusion coefficient (Turing bifurcation threshold) $d_{1[c]}$ such that $C_{min}=0$ is given as 
\begin{equation} \label{eq:3.3}
d_{1[c]}=\frac{d_{2}(a_{11}a_{22}-2a_{12}a_{21})-\sqrt{d_{2}^{2}(a_{11}a_{22}-2a_{12}a_{21})^{2}-d_{2}^{2}a_{11}^{2}a_{22}^{2}}}{a_{22}^{2}}.  
\end{equation}

The system will show stationary and non-stationary patterns for $d_{1}<d_{1[c]}$, but the coexisting equilibrium $(u^{*},v^{*})$ of the local model (\ref{eq:diff1}) remains stable under random heterogeneous perturbation when $d_{1}>d_{1[c]}$.

Moreover, to ensure the positivity of $k^{2}=k^{2}_{\min}$ at the Turing bifurcation threshold, we need to have $d_{2}a_{11}+d_{1}a_{22}>0$, i.e., $d_{1}<d_{2}$ needs to be satisfied for the Turing instability conditions. Hence, the self-diffusion coefficient of the prey population is less than that of the predator population for the model (\ref{eq:diff1}). 

The wavelength at the Turing bifurcation threshold is $\lambda_{m}=2\pi/k_{m}$ where $k_{m}$ is the wavenumber corresponding to the maximum real part of the positive eigenvalue. If the above necessary condition holds and $C_{min}<0$ with certain $k^{2}$ in the interval of $(\zeta^{-}, \zeta^{+})$ where 
\begin{equation} \label{eq:3.4_N}
\begin{aligned}
\zeta^{+}(d_{1})&=\frac{(d_{2}a_{11}+d_{1}a_{22})+\sqrt{(d_{2}a_{11}+d_{1}a_{22})^{2}-4d_{1}d_{2}\det(\textbf{J}(E^*))}}{2d_{1}d_{2}} \\
\zeta^{-}(d_{1})& =\frac{(d_{2}a_{11}+d_{1}a_{22})-\sqrt{(d_{2}a_{11}+d_{1}a_{22})^{2}-4d_{1}d_{2}\det(\textbf{J}(E^*))}}{2d_{1}d_{2}}.
\end{aligned}
\end{equation}
Then $(u^{*},v^{*})$ is an unstable homogeneous steady-state of system (\ref{eq:diff1}). Summarizing the conditions, we get the following theorem:

\begin{theorem} \label{Theorem-3.1}
Considering that the interior equilibrium point $(u^{*},v^{*})$ is locally asymptotically stable, if the following conditions hold
\begin{equation} \label{eq:3.5_N}
(d_{2}a_{11}+d_{1}a_{22})^{2}>4d_{1}d_{2}\det(\textbf{J}(E^*)),\ \  d_{1}<d_{2}
\end{equation}
and there exists a wave-number $k^{2}\in(\zeta^{-},\zeta^{+})$, then the positive constant steady-state solution $(u^{*},v^{*})$ of model (\ref{eq:diff1}) is Turing unstable.
\end{theorem}

The Turing instability analysis for the proposed nonlocal model is almost similar to the analysis corresponding to the local spatio-temporal model (\ref{eq:diff1}). However, we have derived the Turing instability conditions for the nonlocal model (\ref{eq:loc1}) in Appendix \ref{appendixB}.

Local stability analysis for the spatio-temporal model with two-dimensional diffusion will be the same as it is for the case of one-dimensional diffusion. In this case, the perturbation around $(u^{*}, v^{*})$ will take the following form: 
\mathcenter
\begin{align*}
    \begin{pmatrix}
        u \\
        v
    \end{pmatrix}=\begin{pmatrix}
        u^{*} \\
        v^{*}
    \end{pmatrix}+\epsilon \begin{pmatrix}
        u_{1} \\
        v_{1}
    \end{pmatrix}\exp{(\lambda t+i(k_{x}x+k_{y}y))},
\end{align*}
where $|\epsilon|\ll 1$ and $\lambda$ is the growth rate of perturbation. And, $\textbf{k}=(k_{x},k_{y})$ is the wave number vector along with the wave number $k=|\textbf{k}|$.

\section{Numerical Results} \label{sec:5}

This section validates proven analytical findings through numerical simulation. First, the dynamics of the temporal model are explored, and then the significance of incorporating the spatio-temporal model is analyzed. Later, we demonstrated how nonlocal interaction has influenced the dynamic nature of the system. We have adopted the finite difference method to solve the proposed spatio-temporal reaction-diffusion model numerically. This method is a widely used numerical technique for approximating the solutions of partial differential equations. Unless it is mentioned otherwise, we fix some of the parameters used in the model, which are given in Table \ref{Table:1}. 

\begin{table}[!ht]
\centering
\begin{tabular}{|c|c|c|c|c|c|c|c|c|c|c|c|c|c|c|}
\hline
Parameter & $r$ & $\omega$ & $a$ & $m$ & $b$ & $\alpha$ & $\eta$ & $A$ & $d$ & $p$ & $c$ & $l$ & $\mu_{1}$ & $\mu_{2}$ \\
 \hline
Value & $1$ & $10$ & $0.36$ & $0.42$ & $0.5$ & $0.5$ & $0.05$ & $2$ & $0.05$ & $0.05$ & $0.9$ & $0.005$ & $0.1$ & $0.05$  \\
\hline
\end{tabular}
\caption{Parameter values used in the numerical simulations} \label{Table:1}
\end{table}

\subsection{Temporal dynamics}
As we have emphasized the impact of the psychological effect on predator-prey interaction in this work, the fear effect can be considered one of the prime factors in analyzing the system dynamics. Moreover, we have already mentioned that the predators affect each other's activity while hunting and searching for prey. So, a predator-dependent functional response is the most suitable one, and it is considered in this work. 

\begin{figure}[ht!]
     \centering
     \begin{subfigure}{0.4\textwidth}
         \centering
         \includegraphics[width=7.5cm]{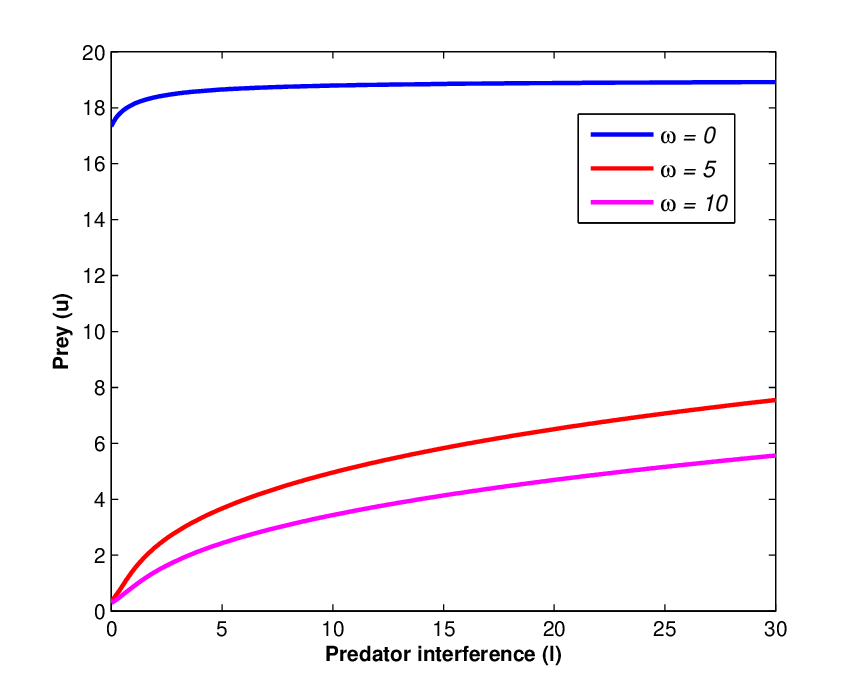}
         \label{fig:1a}
     \end{subfigure}
     \begin{subfigure}{0.4\textwidth}
         \centering
         \includegraphics[width=7.5cm]{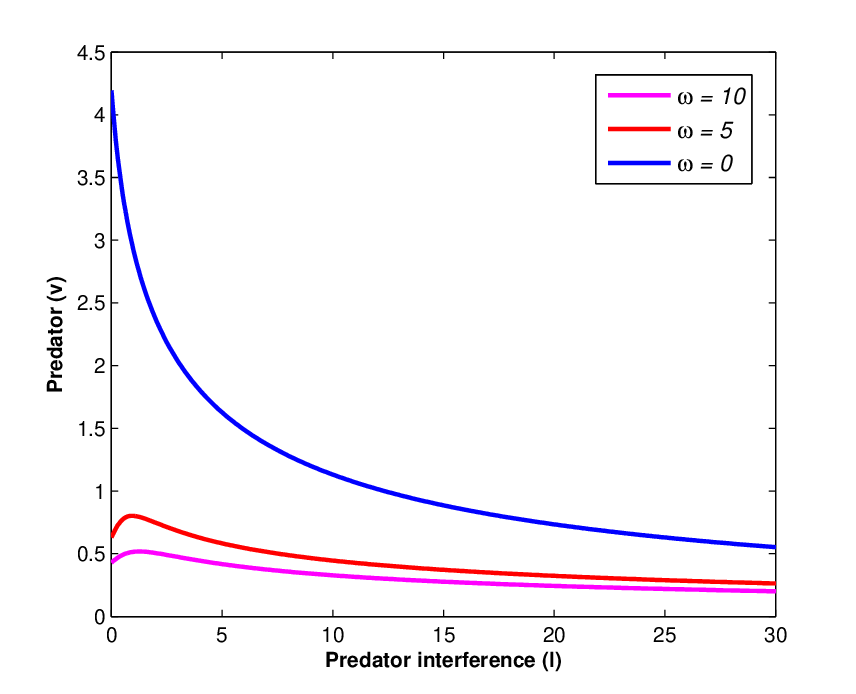}
         \label{fig:1b}
     \end{subfigure}
 \caption{Impact of predator interference on (a) prey (u) and (b) predator (v) population in the presence and absence of the fear effect.} \label{fig:1}
\end{figure}

\begin{figure}[ht!]
     \centering
     \begin{subfigure}{0.4\textwidth}
         \centering
         \includegraphics[width=7.5cm]{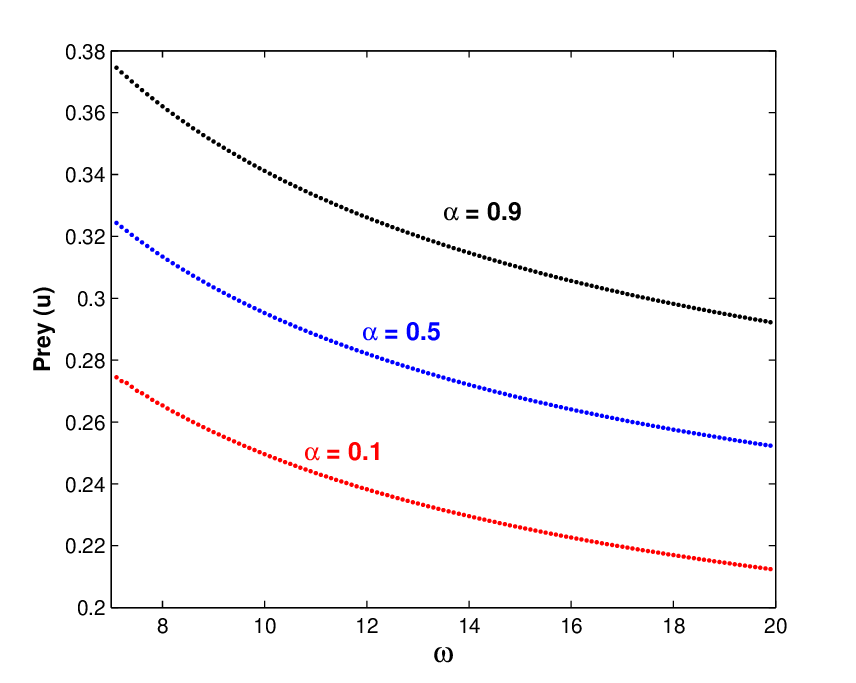}
         \label{fig:2a}
     \end{subfigure}
     \begin{subfigure}{0.4\textwidth}
         \centering
         \includegraphics[width=7.5cm]{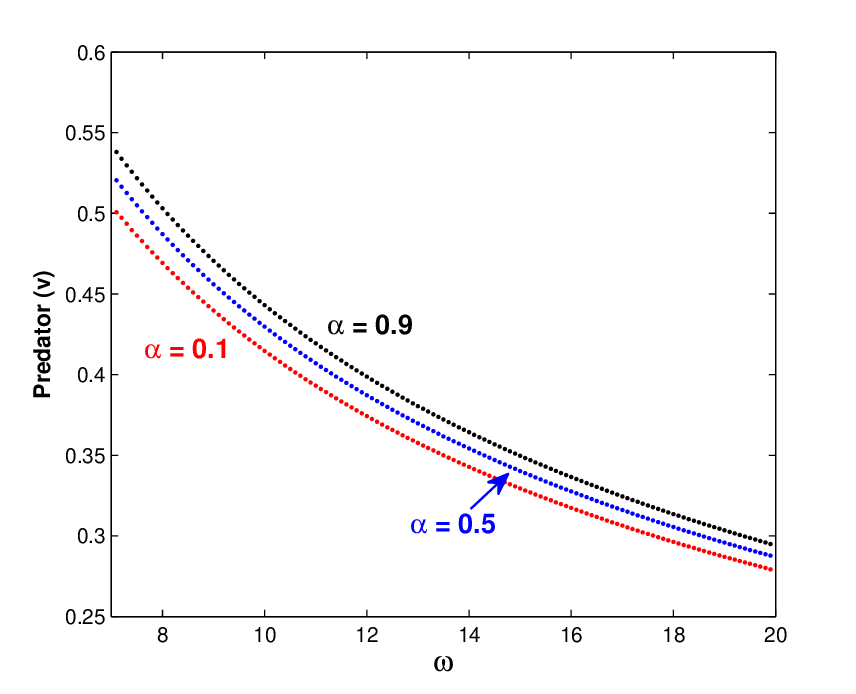}
         \label{fig:2b}
     \end{subfigure}
 \caption{Comparison of the components of $E^{*}$ for $\alpha=0.1,\ 0.5,\ 0.9$ while varying $\omega.$} \label{fig:2}
\end{figure}

Figure \ref{fig:1} depicts how the fear effect reduces the population count in the presence of predator interference $(l)$. It is observed that the more they intrude on each other's business, the more the prey count increases. And this predator interference is more effective when a certain amount of fear exists in the system. Furthermore, the counts of prey $(u)$ and predator $(v)$ species are shown in Fig. \ref{fig:2} for increasing the value of fear level $(\omega)$. In this case, the prey species has shown a declination for increasing value of $\omega$, which causes the reduction in predator population too. But, in this work, we have dealt with the fear effect and provided a source of alternative or additional food to the predators. This figure supports the fact that the inclusion of alternative food helps to increase the prey population as they get a chance to save themselves by hiding in a safe zone, creating prey refuge while predators are busy with secondary food sources, further increasing the predator count. 

\begin{figure}[ht!]
     \centering
     \begin{subfigure}{0.4\textwidth}
         \centering
         \includegraphics[width=7.5cm,height=5.5cm]{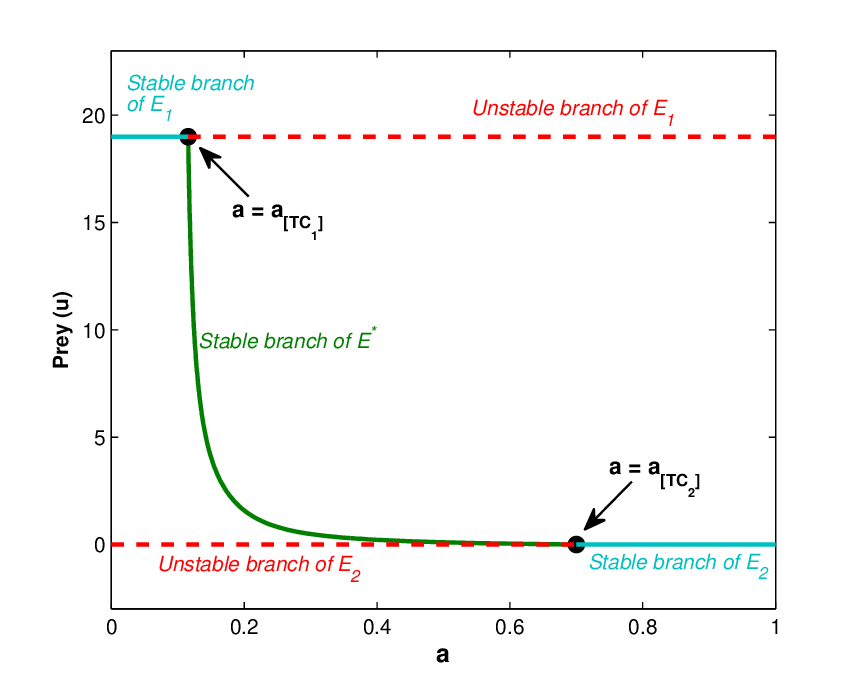}
         \caption{}\label{fig:3a}
     \end{subfigure}
     \begin{subfigure}{0.4\textwidth}
         \centering
         \includegraphics[width=7.5cm,height=5.5cm]{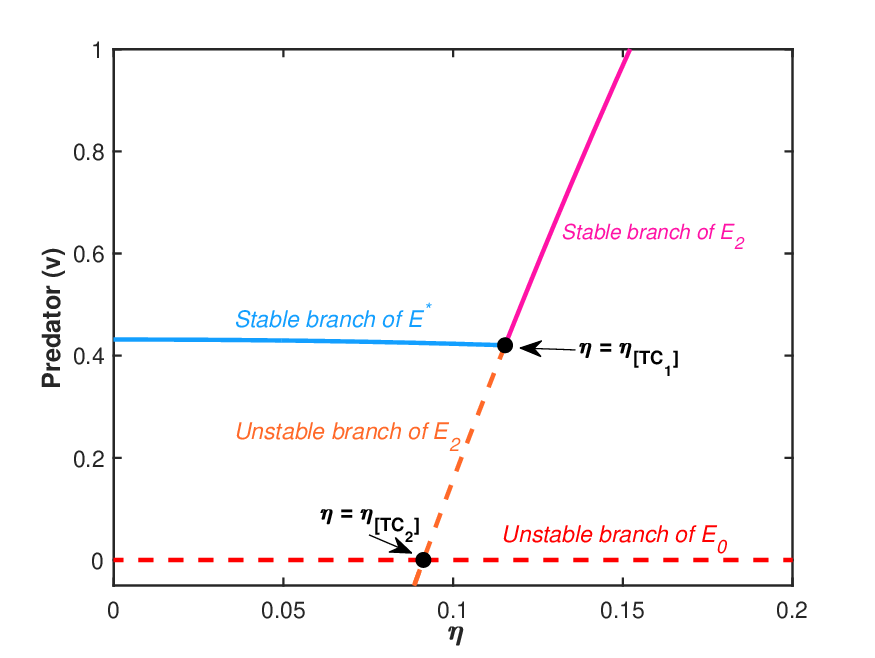}
         \caption{}\label{fig:3b}
     \end{subfigure}
    \begin{subfigure}{0.4\textwidth}
         \centering
         \includegraphics[width=7.5cm,height=5.5cm]{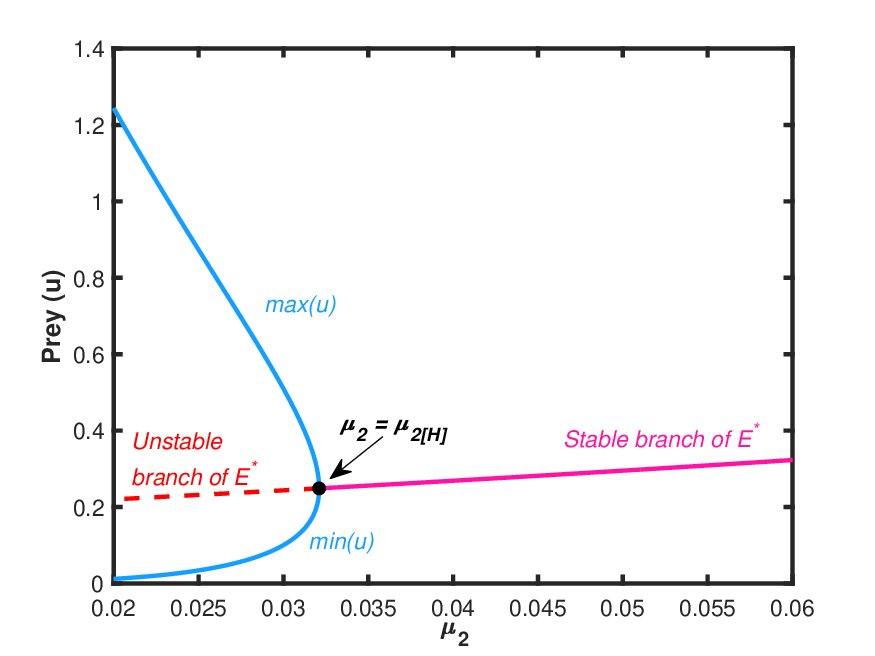}
         \caption{}\label{fig:3c}
     \end{subfigure}
     \begin{subfigure}{0.4\textwidth}
         \centering
         \includegraphics[width=7.5cm,height=5.5cm]{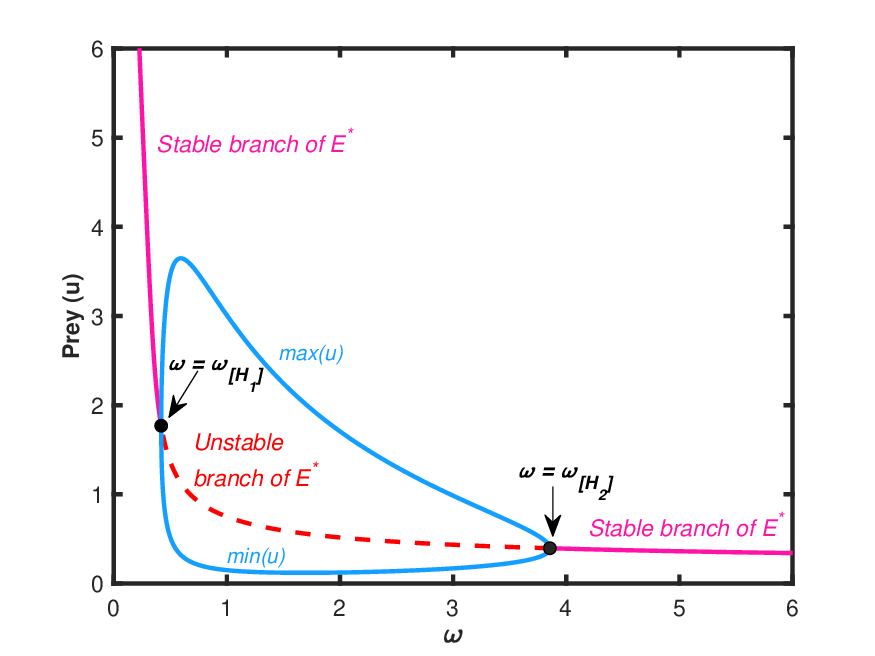}
         \caption{}\label{fig:3d}
     \end{subfigure}
\caption{Change of dynamical behaviour of temporal system (\ref{eq:det1}) with increasing (a) $c$, (b) $\eta$, (c) $\mu_{2}$, and (d) $\omega$.} \label{fig:3}
\end{figure}

We portray some scenarios in Fig. \ref{fig:3} when different system parameters are chosen as control parameters. First, it is observed that the consumption rate of predator $(a)$ contributes to the existence of the steady coexisting state [see Fig. \ref{fig:3a}]. If the predator consumes the targeted prey at a very low rate to their growth, then they may not be able to survive in the system even if there is some additional food present, and a stable predator-free system occurs in such cases (e.g., stable $E_{1}$ in our case). But from this situation, if the consumption rate starts to increase, then a situation arises where both prey and predator exist in a steady state (e.g., stable coexisting equilibrium $E^{*}$). In this case, the coexisting equilibrium point switches the stability behaviour from the predator-free equilibrium $E_{1}$ through transcritical bifurcation. For the considered parameter set, this transcritical bifurcation occurs at $a=a_{[TC_1]}=0.115608$ [see Fig. \ref{fig:3a}]. A further increment of the parameter $a$ leads to the extinction of prey species, and only predator species exist in a steady state in such cases (e.g., stable prey-free equilibrium $E_{2}$). In this case, the coexisting equilibrium point and prey-free equilibrium point exchange their stabilities through another transcritical bifurcation that occurs at $a=a_{[TC_2]}=0.699983$ [see Fig. \ref{fig:3a}]. 

We have studied the model's behaviour when the predators go for additional food sources instead of the targeted prey, and the parameter $\eta$ is involved in signifying it. Figure \ref{fig:3b} shows that if the predator detects the additional food up to a certain amount along with the prey species, then both the population exist as a steady state, but increasing the parametric value leads to a situation where the two equilibrium points $E_{2}$ and $E^{*}$ exchange their stability through a transcritical bifurcation which occurs at $\eta = \eta_{[TC_{1}]}=0.115306$. The prey-free equilibrium exists for $\eta>\eta_{[TC_{1}]}$. Also, there exists an unstable branch of $E_{2}$ when $\eta$ lies below $\eta_{[TC_{1}]}$, which ultimately emerges with the unstable trivial equilibrium $E_{0}$ through another transcritical bifurcation at $\eta = \eta_{[TC_{2}]}=0.091241$. 

Now, we see how the intra-specific mortality rate of predators regulates the system dynamics, which can be controlled by the parameter $\mu_{2}$. When a predator is exposed to prey species that are inadequate in the system, they tend to compete with each other to gain more food. This psychology remains the same even in the presence of alternative food sources. Figure \ref{fig:3c} shows that a certain amount of competition is needed for the coexisting equilibrium point. But both the populations start to oscillate around $E^{*}$ if $\mu_{2}$ comes below a certain threshold value through Hopf bifurcation, which occurs at $\mu_{2[H]}=0.032103$. In this case, the temporal model exhibits a supercritical Hopf bifurcation, and the first Lyapunov coefficient is $l_{1}=-0.0294$ \cite{perko2013differential}. Furthermore, it is observed that the fear term $(\omega)$ acts as a stabilizing as well as destabilizing factor in the system [see Fig. \ref{fig:3d}]. Both the populations coexist and are stable for a small value of $\omega$, but an increase in the value leads to oscillation, indicating the occurrence of a supercritical Hopf bifurcation at $\omega_{[H_1]}=0.419521$ with the first Lyapunov coefficient $l_{1}=-0.0162$. A stable limit cycle is generated through this Hopf bifurcation and vanishes through another supercritical Hopf bifurcation at $\omega_{[H_2]}=3.855245$ with $l_{1}=-0.0305$. With further increased fear parameter value $\omega$, the coexisting equilibrium point remains stable [see Fig. \ref{fig:3d}].

\subsection{Effects of prey refuge $(m)$ and additional food $(A)$ on spatio-temporal interactions}
\begin{figure}[htb!]
     \centering
     \begin{subfigure}[t]{0.4\textwidth}
         \centering
         \includegraphics[width=7.5cm,height=5.5cm]{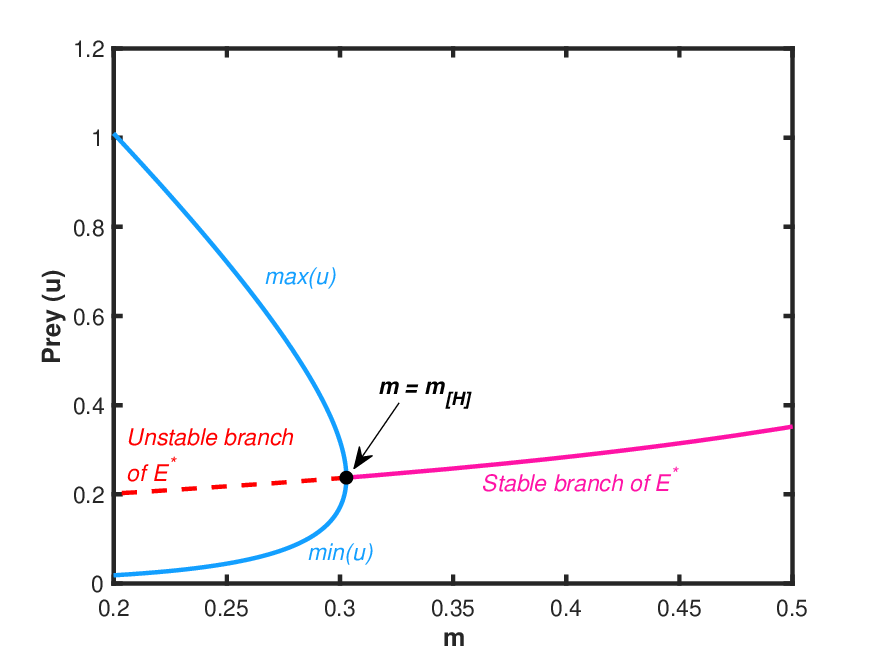}
         \caption{}\label{fig:4a}
     \end{subfigure}
     \begin{subfigure}[t]{0.4\textwidth}
         \centering
         \includegraphics[width=7.7cm,height=5.5cm]{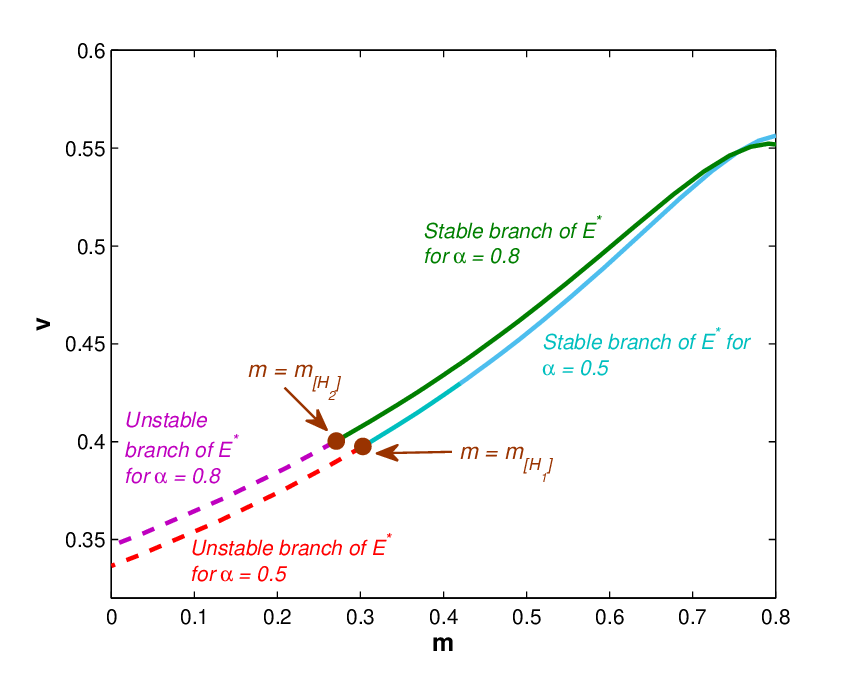}
         \caption{}\label{fig:4b}
     \end{subfigure}
\caption{(a) Change of dynamical behaviour of the temporal system with increasing $m$. (b) Impact of prey refuge $(m)$ on the predator population $(v)$ in the presence of alternative food sources. } \label{fig:4}
\end{figure}

Sometimes, a predator fails to access the whole of the prey population when a prey species successfully hides in a safe zone to dodge the frequent attack of their predator. The prey refuge parameter $(m)$ in the model can capture this scenario. Figure \ref{fig:4a} shows that the coexisting equilibrium point can be obtained when a certain amount of prey hides in a predator-prohibited zone, but both populations perform oscillatory behaviour if the prey refuge parameter comes below a threshold value. 
This situation occurs in the system through a supercritical Hopf bifurcation around the coexisting equilibrium point $(E^{*})$ at $m_{[H]}=0.302811$ (the first Lyapunov coefficient is $l_{1}=-0.0389$), and a stable limit cycle exists for $m<m_{[H]}$. Furthermore, the predator count has been plotted for increasing prey refuge $(m)$ in Fig. \ref{fig:4b} for different amounts of provided additional food sources. It is seen that the Hopf threshold shifts towards the left due to the incorporation of more additional food. Therefore, the additional food expands the scope of population coexistence as the population oscillates when $m$ is less than the Hopf threshold [see Fig. \ref{fig:4a}]. 
\begin{figure}[htb!]
     \centering
         \includegraphics[width=8cm]{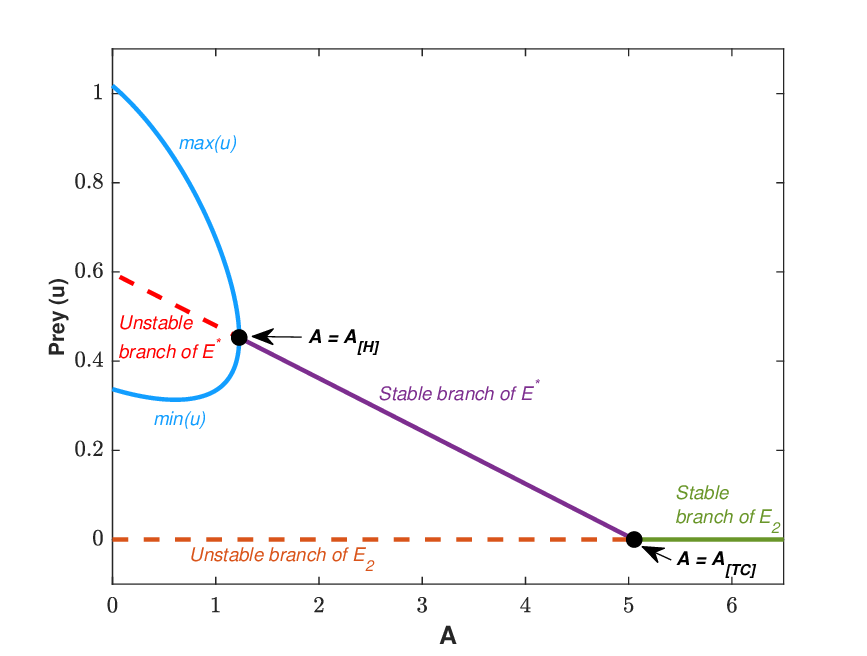}
\caption{Change of dynamical behaviour of the temporal system with increasing $A$. The parametric values are chosen from Table \ref{Table:1} and $\omega=5$.} \label{fig:sp1}
\end{figure}
In the proposed model, the parameter $A$ represents the additional food provided to the predator species, which is considered a bifurcation parameter in Fig. \ref{fig:sp1}. It is observed that when a small amount of additional food is introduced, the system exhibits oscillatory behaviour. However, as it increases, the system stabilizes into a coexistence equilibrium. This stability switching occurs through a supercritical Hopf bifurcation around $E^{*}$ at $A=A_{[H]}=1.227$, exceeding which a stable interior equilibrium emerges. Nevertheless, this stable coexistence does not persist for a long range of $A$. With a further increase in the parameter value, the predator population continues to grow while the prey biomass declines. Eventually, the prey population diminishes to the extent that it merges with the prey-free equilibrium via a transcritical bifurcation at $A=A_{[TC]}=5.053$. Beyond this threshold, the system stabilizes in a predator-only equilibrium, where the prey species is absent.
\begin{figure}[htb!]
     \centering
     \begin{subfigure}[t]{0.4\textwidth}
        \centering
         \includegraphics[width=\linewidth]{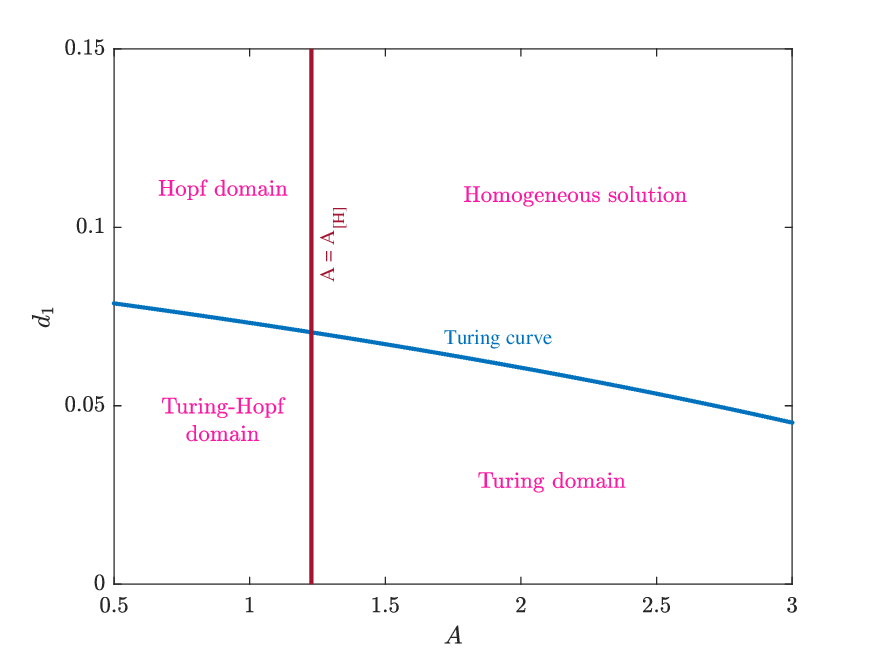}
         \caption{}\label{fig:sp2a}
     \end{subfigure}
     ~
     \begin{subfigure}[t]{0.4\textwidth}
        \centering
         \includegraphics[width=\linewidth]{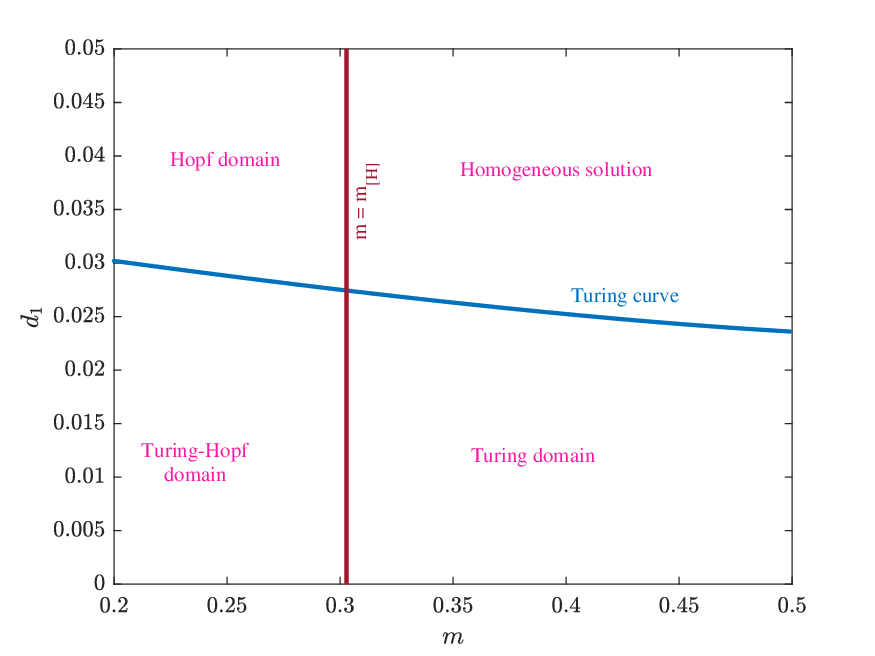}
         \caption{}\label{fig:sp2b}
     \end{subfigure}
\caption{Temporal-Hopf ({\color{darkmar}\solidrule}) and Turing bifurcation curve ({\color{fadblu}\solidrule}) in the (a) $A$-$d_{1}$ plane and (a) $m$-$d_{1}$ plane. For the left panel, $\omega$ is chosen as $5$, while for the right one, it is considered $10$. The other parametric values are chosen from Table \ref{Table:1}.} \label{fig:sp2}
\end{figure}

Turing instability is a key phenomenon studied in reaction-diffusion models, facilitating the emergence of non-homogeneous stationary patterns. To identify Turing instability, the coexisting homogeneous steady state must be locally asymptotically stable. In the proposed temporal model, a stable coexistence equilibrium is observed only when either the prey refuge parameter $(m)$ or the additional food parameter $(A$) exceeds their respective Hopf bifurcation thresholds, denoted as $m_{[H]}$ and $A_{[H]}$, respectively [see Figs. \ref{fig:4a} and \ref{fig:sp1}]. Now, Fig. \ref{fig:sp2} illustrates the Turing bifurcation curves for varying levels of prey refuge $(m)$ and additional food $(A)$, demonstrating that the critical prey diffusion threshold $(d_{1[c]})$ decreases as these parameters increase. Stationary Turing patterns emerge when $d_{1}<d_{1[c]}$, provided that the chosen parameters lie within the Hopf stable domain. However, when the parameters are chosen from the Hopf unstable domain, the system exhibits either non-homogeneous stationary patterns or oscillatory solutions, depending on the position of $d_{1}$ relative to the Turing curve. Specifically, non-homogeneous stationary patterns arise when $d_{1}$ is below the Turing curve, whereas oscillatory solutions occur when it lies above it. So, Fig. \ref{fig:sp2} indicates that an increase in either prey refuge or additional food reduces the spatial extent of species colonization by narrowing down the Turing domain. This suggests that higher values of these parameters lower the prey diffusion threshold $(d_{1[c]})$ required for the emergence of Turing patterns, thereby restricting the parameter space favourable for patch formation.

\subsection{Impact of fear effect through local and nonlocal interactions}
Now, we analyze the results by incorporating the fear effect through local interactions in the temporal model. This work focuses on studying the impact of fear in the presence of additional food sources. So, we choose the fear level $\omega$ as one of the bifurcation parameters. In our temporal model, there exists two Hopf bifurcation thresholds $\omega_{[H_1]}$ and $\omega_{[H_2]}$. The stable coexisting equilibrium point is found when $\omega$ lies outside these thresholds, and the system shows periodic dynamics inside these thresholds [see Fig. \ref{fig:3d}]. We first focus on the temporal Hopf stable domain $\omega>\omega_{[H_2]}$.

\begin{figure}[htb!]
     \centering
     \begin{subfigure}{0.4\textwidth}
         \centering
         \includegraphics[width=7cm]{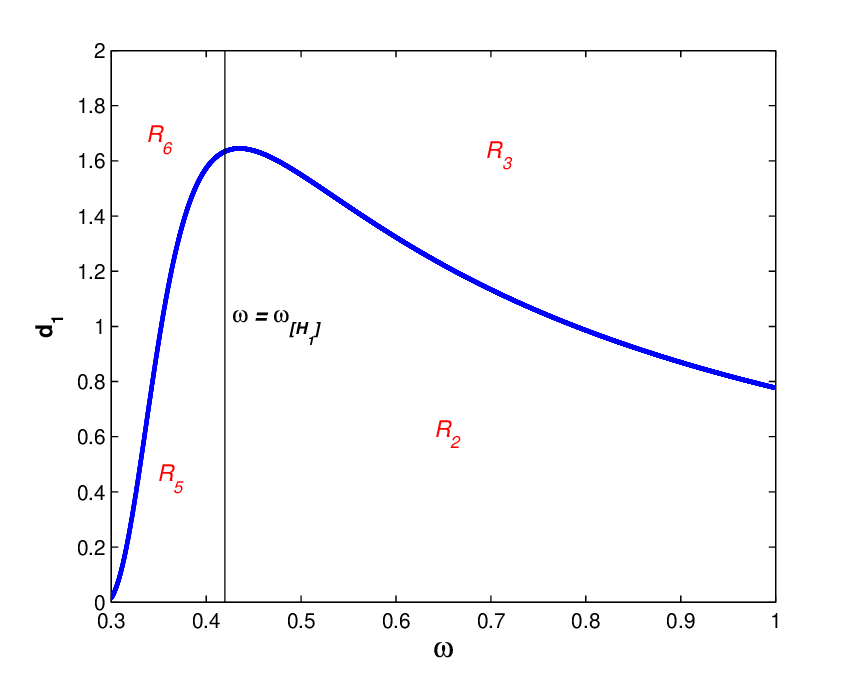}
         \caption{}\label{fig:5a}
     \end{subfigure}
     \begin{subfigure}{0.4\textwidth}
         \centering
         \includegraphics[width=7cm]{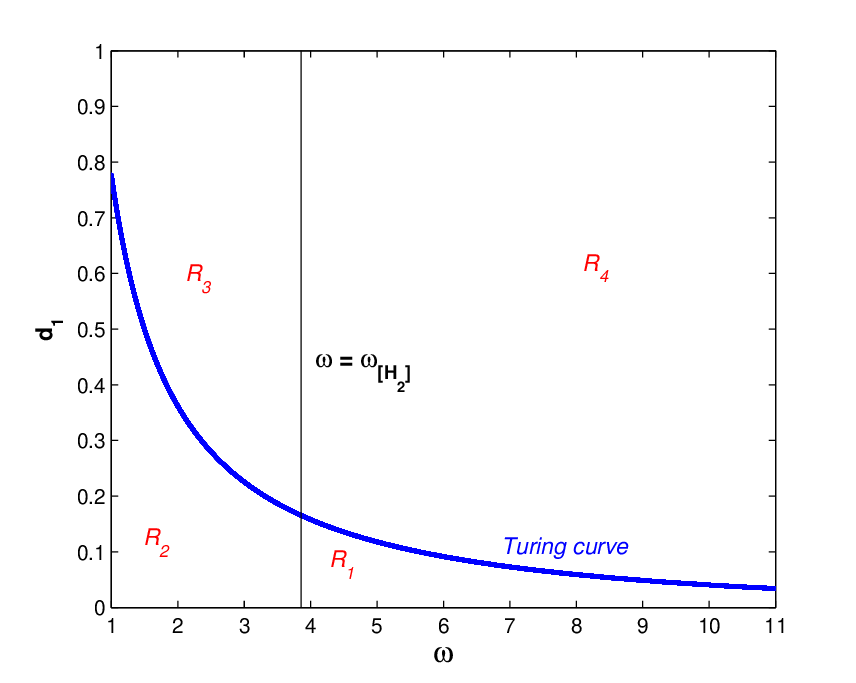}
         \caption{}\label{fig:5b}
     \end{subfigure}
\caption{Temporal-Hopf, Turing bifurcation curves in the $\omega$-$d_{1}$ plane for the local model are represented by black and blue colors. The Turing domain corresponding to (a) $\omega<\omega_{[H_1]}$ and (b) $\omega>\omega_{[H_2]}$.} \label{fig:5}
\end{figure}

For a fixed $\omega$, we obtain the Turing bifurcation threshold $d_{1[c]}$, and we plot this set of Turing bifurcation thresholds for a set of parameter values of $\omega$ in the $\omega$-$d_{1}$ plane, which is shown in Figure \ref{fig:5}. We have plotted the Turing curve and temporal Hopf curve here, which intersect each other, and they divided the region into four sub-regions. Pure Turing domain $(R_{1})$ and homogeneous solution $(R_{4})$ lie on the right of the second Hopf curve $\omega=\omega_{[H_2]}$, below and above the Turing curve, respectively. On the other hand, there is another Turing domain $(R_{5})$ and homogeneous solution $(R_{6})$ exist left to the first Hopf curve $\omega=\omega_{[H_1]}$, below and above Turing curve respectively. The bottom region $(R_{2})$ between the two temporal Hopf curves is the Turing-Hopf domain, while the upper region $(R_{3})$ is the Hopf domain. Here, we have mainly chosen the values of $\omega$ and $d_{1}$ from $R_{1}, R_{2}, R_{3}$ and $R_{4}$ domains as the other two regions $R_{5}$ and $R_{6}$ will show the same dynamical nature as $R_{1}$ and $R_{4}$ respectively. 

\begin{figure}[ht!]
\centering
\begin{subfigure}{0.3\textwidth}
    \includegraphics[width=6cm,height=5cm]{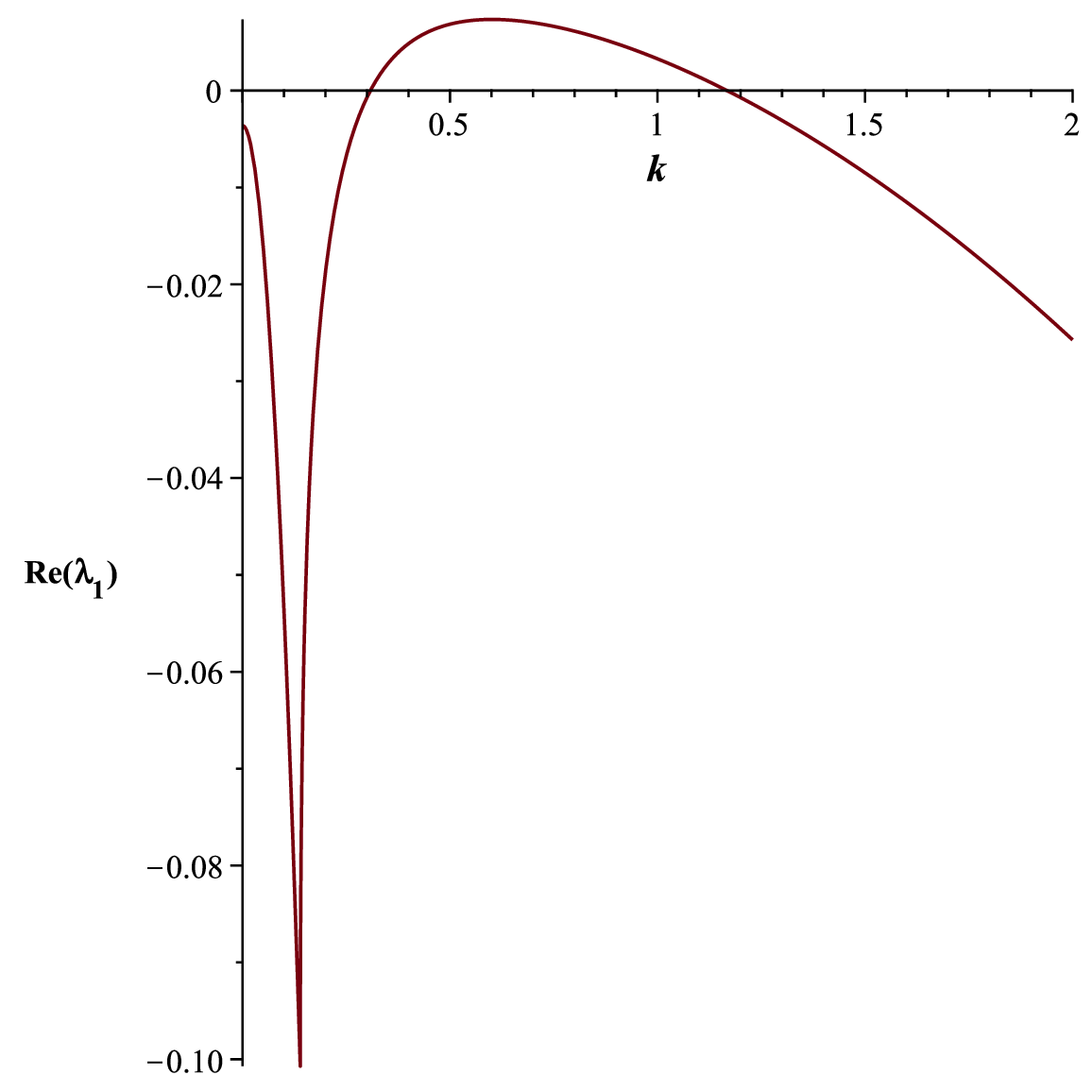}
    \caption{}\label{fig:6a}
\end{subfigure}
\begin{subfigure}{0.3\textwidth}
    \includegraphics[width=6cm,height=5cm]{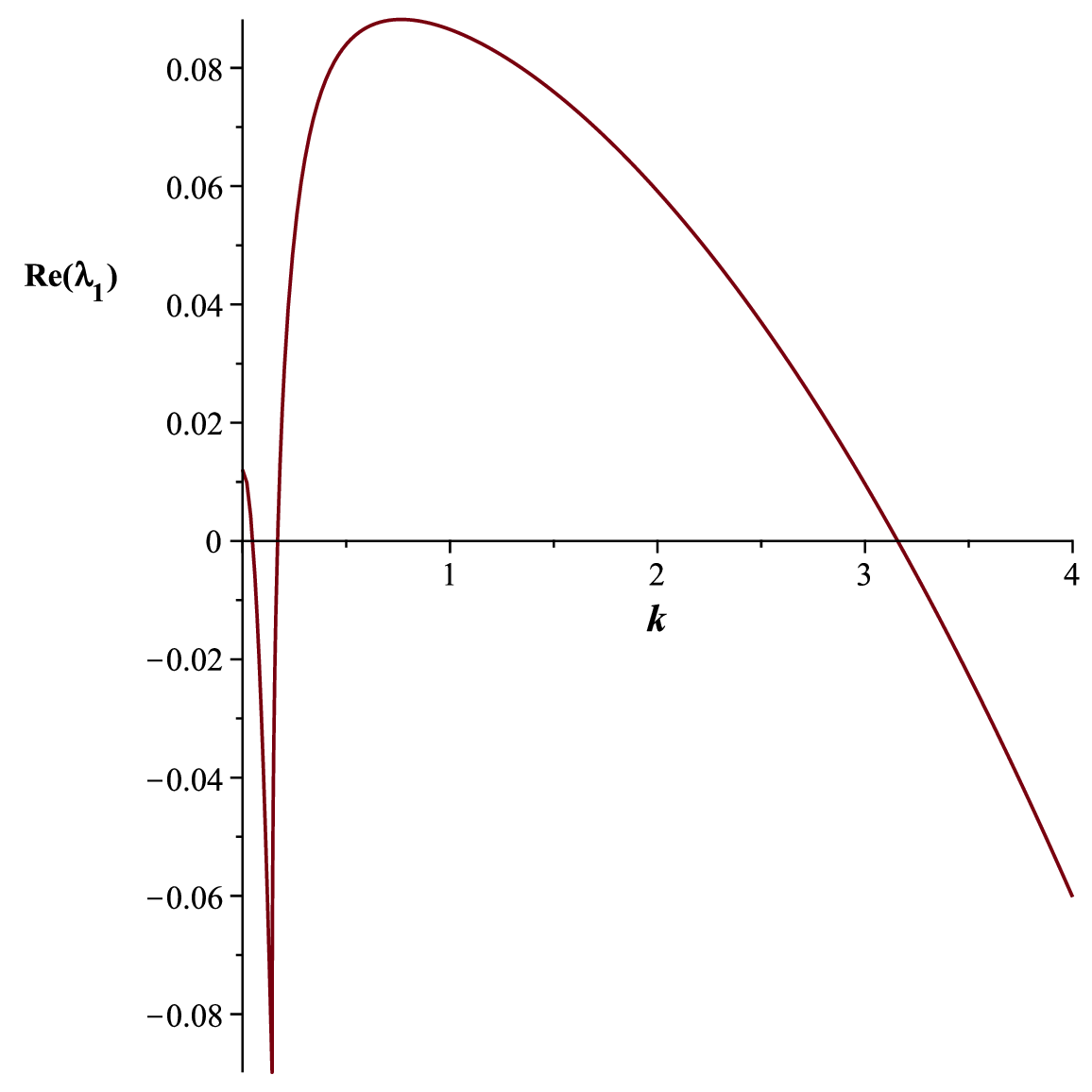}
    \caption{}\label{fig:6b}
\end{subfigure}
\begin{subfigure}{0.3\textwidth}
    \includegraphics[width=6cm,height=5cm]{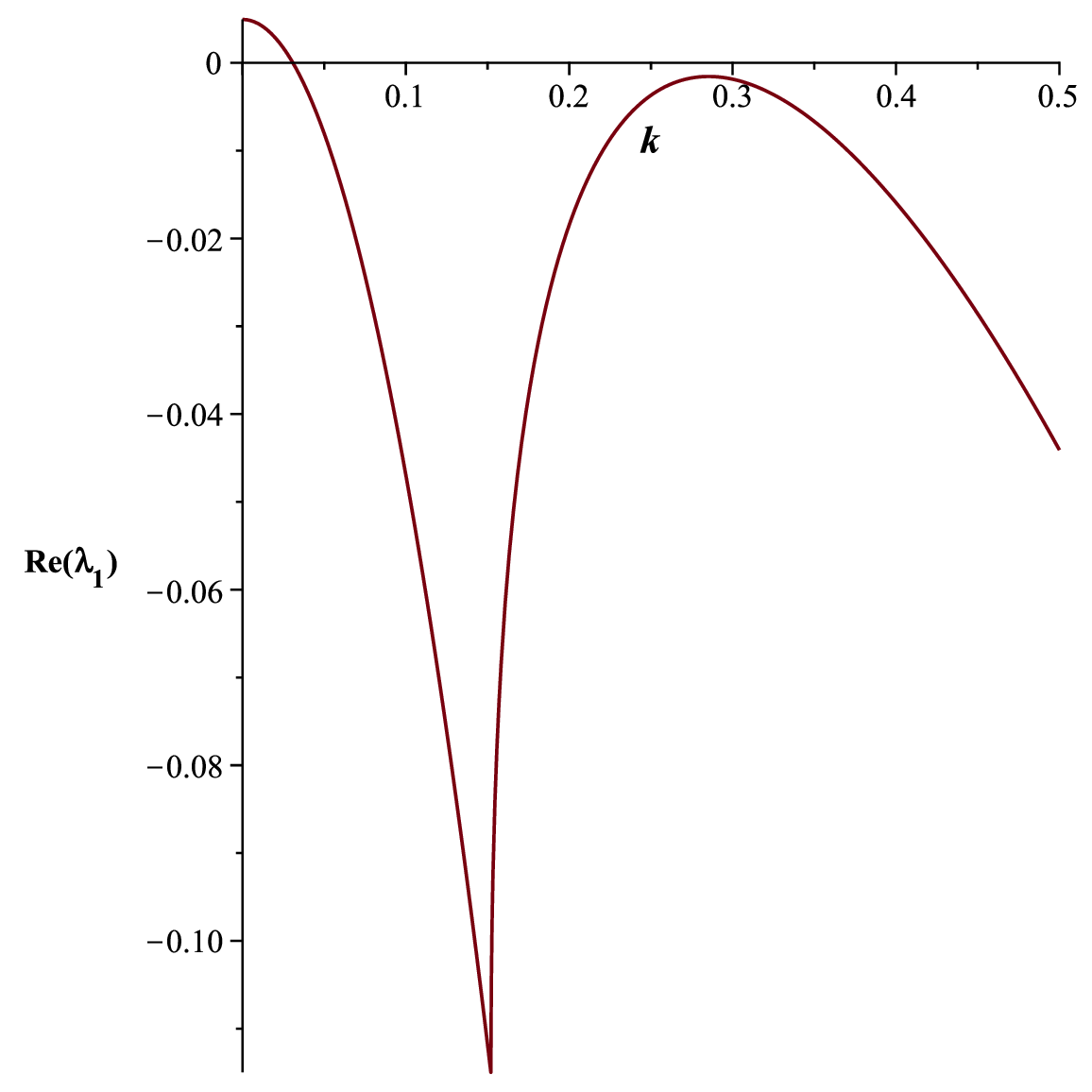}
    \caption{}\label{fig:6c}
\end{subfigure}
\caption{Plots of largest real part of eigenvalues with respect to $k$ when $(\omega,\ d_{1})$ are chosen from $R_{1}, R_{2}$ and $R_{3}$ respectively. (a) Turing domain: $(\omega,\ d_{1})=(10,0.01)$, (b) Turing-Hopf domain: $(\omega,\ d_{1})=(1,0.01)$ and (c) Hopf domain: $(\omega,\ d_{1})=(2,0.38)$.}\label{fig:6}
\end{figure}

\begin{figure}[ht!]
     \centering
     \begin{subfigure}{0.4\textwidth}
         \centering
         \includegraphics[width=7.5cm]{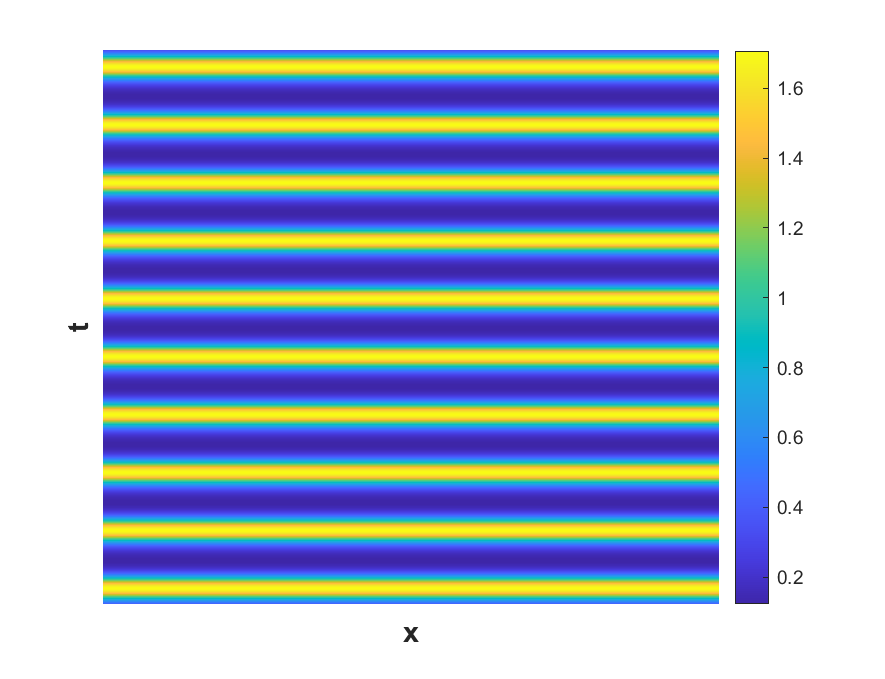}
     \end{subfigure}
     \begin{subfigure}{0.4\textwidth}
         \centering
         \includegraphics[width=7.5cm]{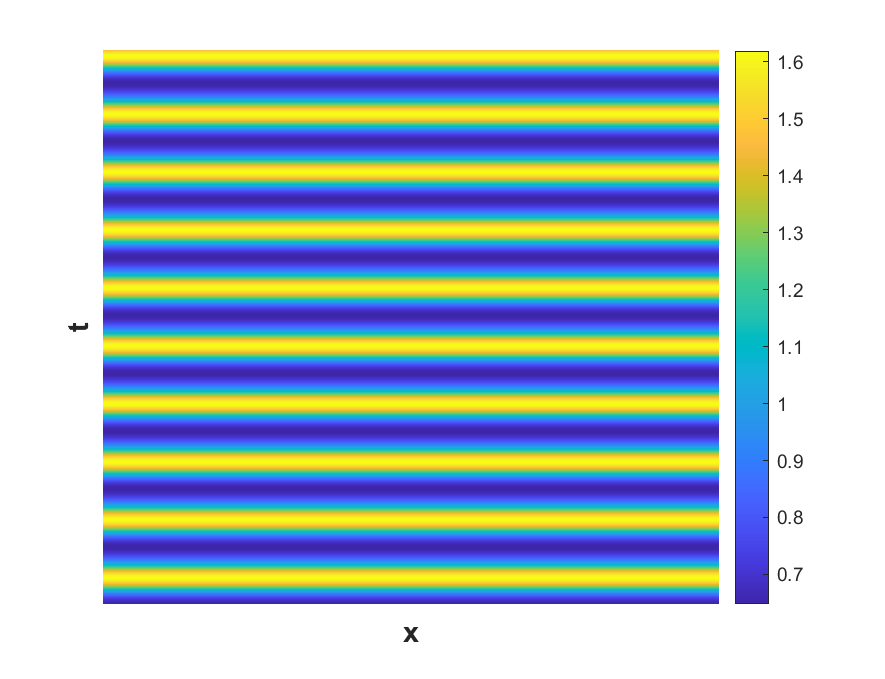}
     \end{subfigure}
 \caption{Contour plots of $u$ (left) and $v$ (right) of the local model for $\omega=2$ when $d_{1} =0.38$ is chosen from Hopf domain $(R_{3}$).} \label{fig:8}
\end{figure}

\begin{figure}[ht!]
\centering
\begin{subfigure}{0.4\textwidth}
\centering
\includegraphics[width=7.5cm, height=4.85cm]{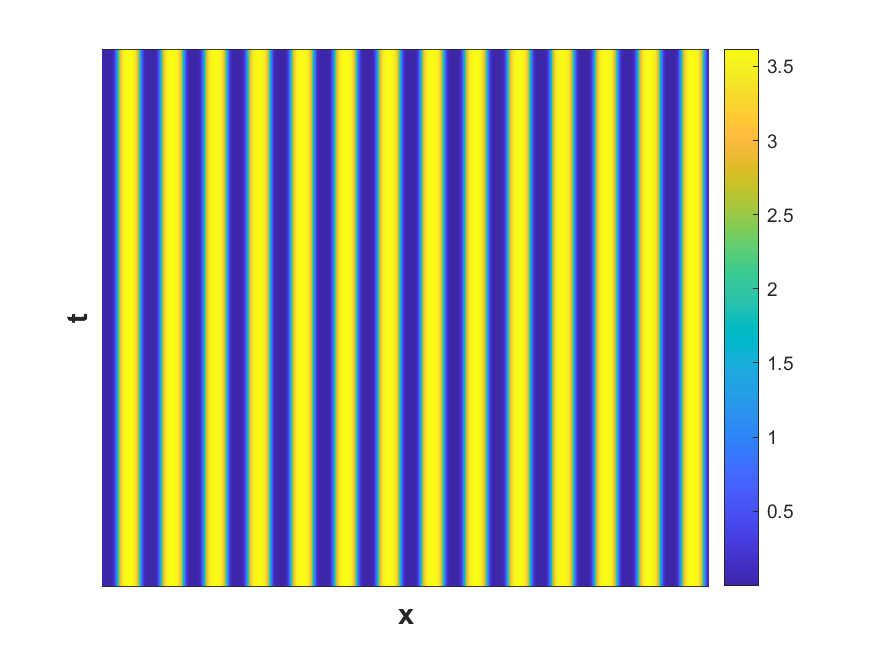}
\end{subfigure}
\begin{subfigure}{0.4\textwidth}
\centering
\includegraphics[width=7.5cm]{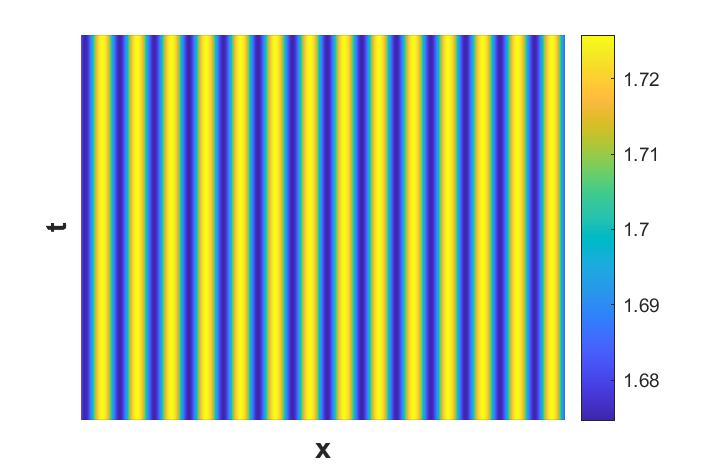}
\end{subfigure}
\caption{Contour plots of $u$ (left) and $v$ (right) of system (\ref{eq:diff1}) for $\omega=1$ when $d_{1}(=0.01)$ is chosen from Turing-Hopf domain $(R_{2}$).}\label{fig:10}
\end{figure}

\begin{figure}[ht!]
     \centering
     \begin{subfigure}{0.4\textwidth}
         \centering
         \includegraphics[width=7.5cm]{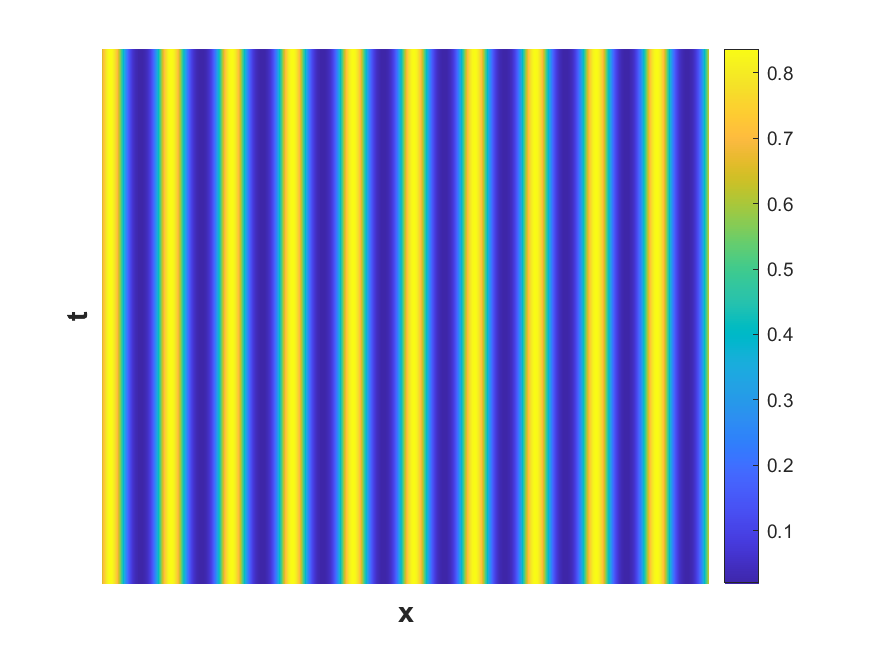}
     \end{subfigure}
     \begin{subfigure}{0.4\textwidth}
         \centering
         \includegraphics[width=7.5cm]{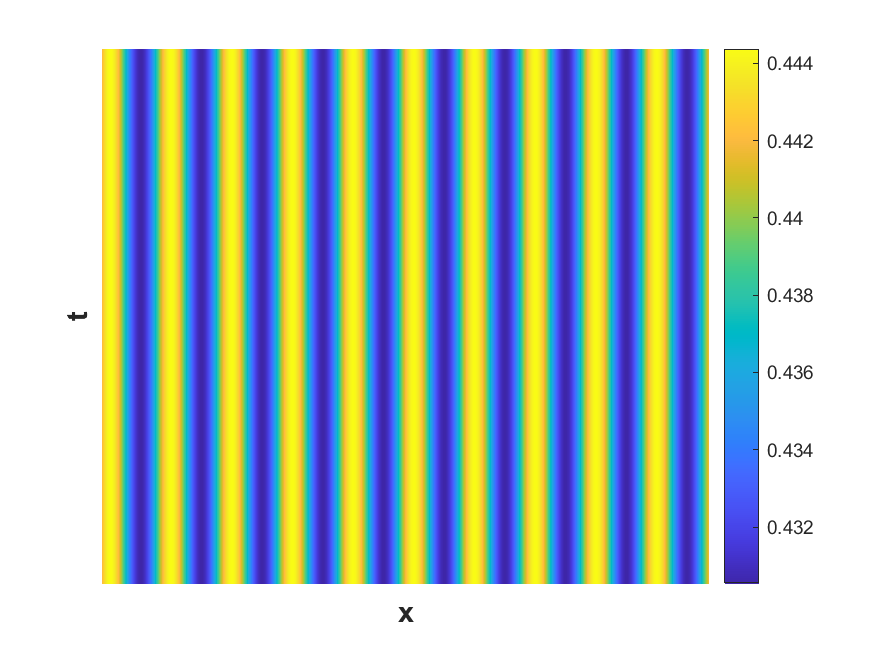}
     \end{subfigure}
\caption{Contour plots of $u$ (left) and $v$ (right) of system (\ref{eq:diff1}) for $\omega=10$ when $d_{1}(=0.01)$ is chosen from Turing domain $(R_{1}$).} \label{fig:11}
\end{figure}

To describe the Turing and non-Turing patterns for the system (\ref{eq:diff1}), we have chosen the spatial domain as $[-L, L]$, where $L=50$, with non-negative initial and periodic boundary conditions. A heterogeneous perturbation is given around the coexisting homogeneous steady-state as the initial conditions to observe the dynamics. We choose small amplitude random perturbations given by $u(x_{j},0)=u^{*}+\epsilon\xi_{j}$ and $v(x_{j},0)=v^{*}+\epsilon\psi_{j}$ with $\epsilon=10^{-5}$ and $\xi_{j}$ and $\psi_{j}$ are Gaussian white noise $\delta$-correlated in space. The dynamical behaviour of the proposed spatio-temporal model is explored in Figs. \ref{fig:5}--\ref{fig:11}. It is also noted that the nonlocal model (\ref{eq:loc1}) turns into a local model (\ref{eq:diff1}) if the range of nonlocal interaction $\delta$ tends to $0$. 

For $\omega=10$ the temporal model (\ref{eq:diff1}) has the feasible interior equilibrium point $E^{*} = (0.295, 0.430)$, which is stable. Also, from equation (\ref{eq:3.3}), $d_{1[c]}$ is found to be $0.0406$ when $d_{2}=10$. The real part of an eigenvalue is plotted in Fig. \ref{fig:6} when $d_{1}$ is chosen from Turing domain $(R_{1})$ [see Fig. \ref{fig:6a}], Turing-Hopf domain $(R_{2})$ [see Fig. \ref{fig:6b}] and Hopf domain $(R_{3})$ [see Fig. \ref{fig:6c}]. Here, the real parts match the cases for the temporal model for $k=0$. The oscillatory solution also occurs for the local model in the domain above the Turing curve and to the left of the temporal Hopf curve. These oscillatory solutions are homogeneous in space but periodic in time. A sample solution is plotted in Fig. \ref{fig:8} for $\omega = 2$ and $d_{1} = 0.38$. 

Some literature already states that if a specialist predator is considered in a predator-prey model, then time-dependent spatial patterns occur when the diffusion parameter is chosen from a bit inside of the temporal Hopf domain \cite{petrovskii2001wave, petrovskii2003quantification}. The Turing behaviour mainly dominates and creates stationary patterns in the Turing-Hopf domain. However, the Hopf behaviour also dominates and produces oscillatory patterns. These oscillatory solutions can be found in this domain in a small region near the Turing curve. Generally, non-homogeneous stationary patterns exist in most parts of the Turing-Hopf domain for the local model, depicted in Fig. \ref{fig:10} for $d_{1}=0.01$ and with $\omega=1$.  

When $d_{1}>d_{1[c]}$ holds in the right of the temporal Hopf curve, the stationary homogeneous solution can be obtained in the stable domain $(R_{4})$. However, a decrease of the diffusion parameter $d_{1}$ makes a shift in the Turing domain $(R_{1})$, where Turing pattern solutions can be obtained for the mentioned boundary condition. In order to draw Fig. \ref{fig:11}, let us choose $d_{1}=0.01$ from the Turing domain (see $R_{1}$ in Fig. \ref{fig:5}) when $\omega=10$. The figure depicts the stationary Turing patterns for the periodic boundary condition when parameter values are chosen from Table \ref{Table:1} along with $d_{2}=10$. The figure shows that the amplitude of the pattern is constant in the whole domain. 

\begin{figure}[ht!]
     \centering
     \begin{subfigure}{0.4\textwidth}
         \centering
         \includegraphics[width=7.5cm]{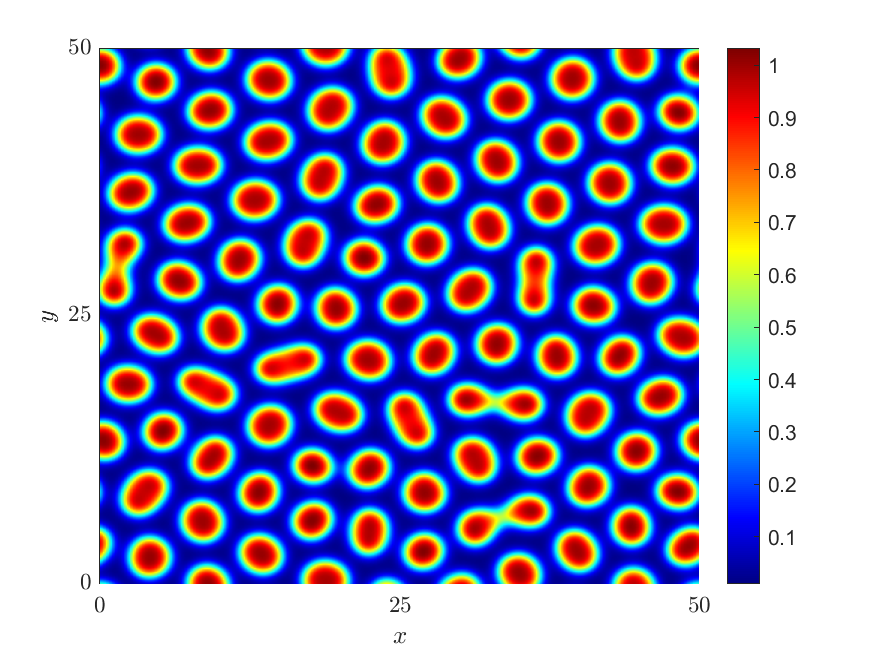}
         \caption{$(\omega, d_{1})=(10,0.01)$ in $R_1$}\label{fig:11_2a}
     \end{subfigure}
     \begin{subfigure}{0.4\textwidth}
         \centering
         \includegraphics[width=7.5cm]{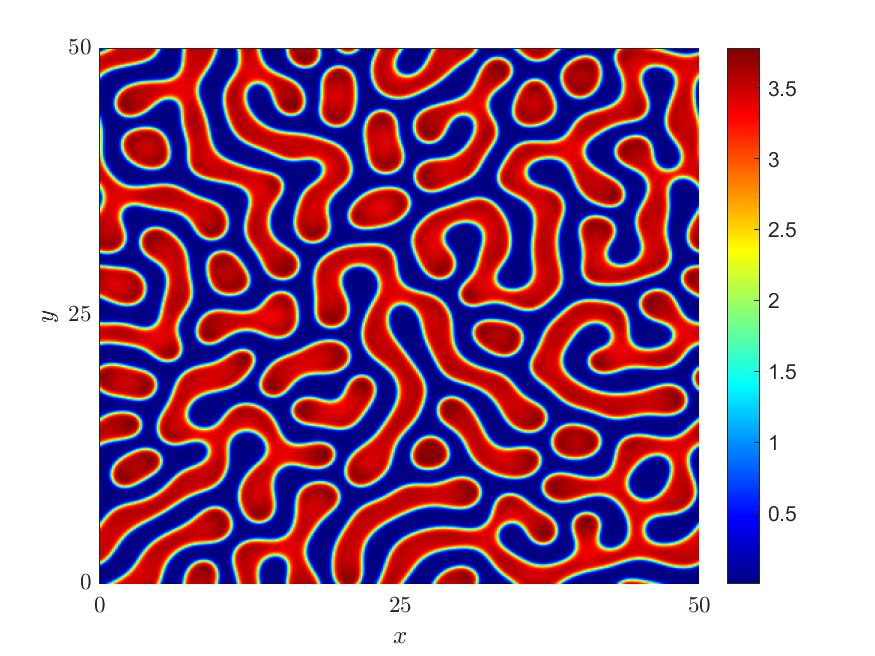}
         \caption{$(\omega, d_{1})=(1,0.01)$ in $R_2$}\label{fig:11_2b}
     \end{subfigure}
     \\
     \begin{subfigure}{0.4\textwidth}
         \centering
         \includegraphics[height=5.5cm,width=7.5cm]{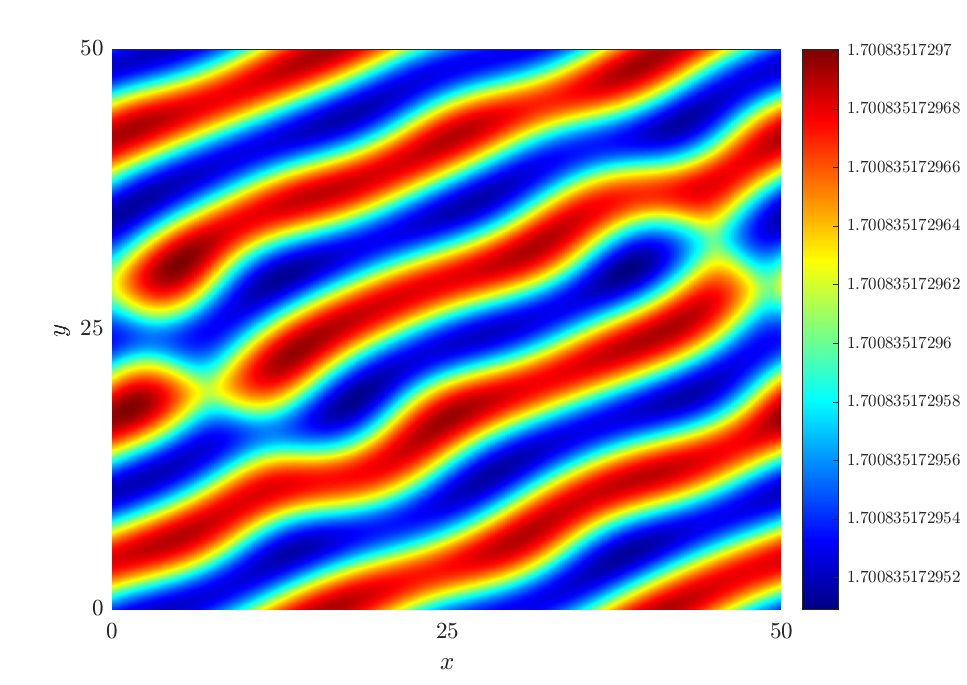}
         \caption{$(\omega, d_{1})=(2,0.38)$ in $R_3$}\label{fig:11_2d}
     \end{subfigure}
     \begin{subfigure}{0.4\textwidth}
         \centering
         \includegraphics[width=7.5cm]{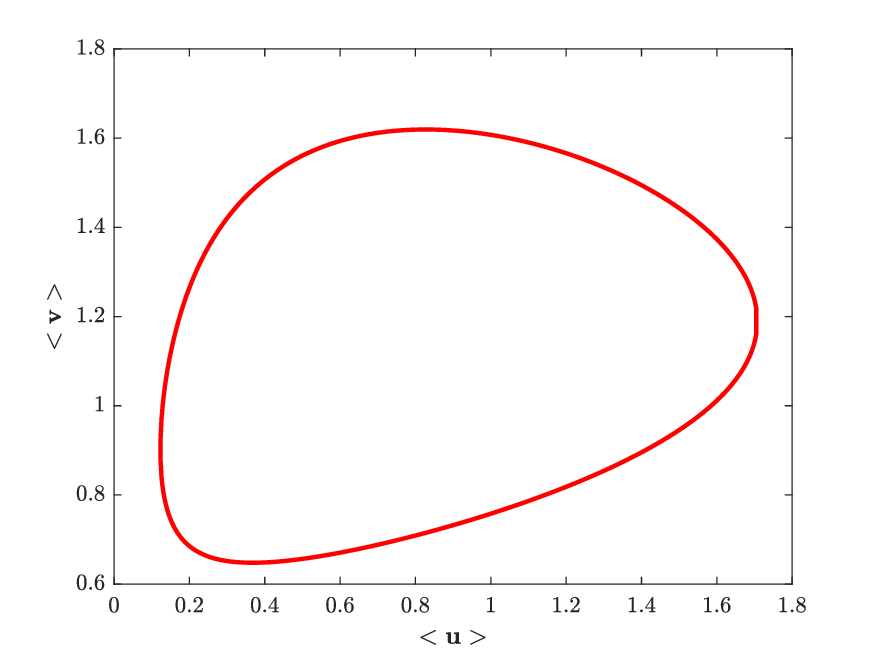}
         \caption{Spatial average of $u$ and $v$ for $t\in [3000, 3500 ]$.}\label{fig:11_2c}
     \end{subfigure}
\caption{Solutions for the prey species in two dimensions for different values of $(\omega, d_{1})$. The spatial domain is chosen as $[0,50]\times[0,50]$ with $dx=dy=0.25$ and $dt=0.0005$.} \label{fig:11_2}
\end{figure}

In Fig. \ref{fig:11_2}, the spatio-temporal patterns have been portrayed when the species disperse in a two-dimensional domain and $(\omega, d_{1})$ are chosen from Turing, Turing-Hopf and Hopf domains. It is observed that the model shows a hot spot pattern when the parameters lie in the Turing domain. In addition, a labyrinthine pattern is produced when the parameter is chosen from the Turing-Hopf domain. Furthermore, the phase portrait of the spatio-temporal model in $\left<u\right>$-$\left<v\right>$ plane is also shown when $\omega$ is considered from Hopf unstable domain but $d_{1}>d_{1[c]}$.

\begin{figure}[ht!]
\centering
\includegraphics[width=11cm,height=7cm]{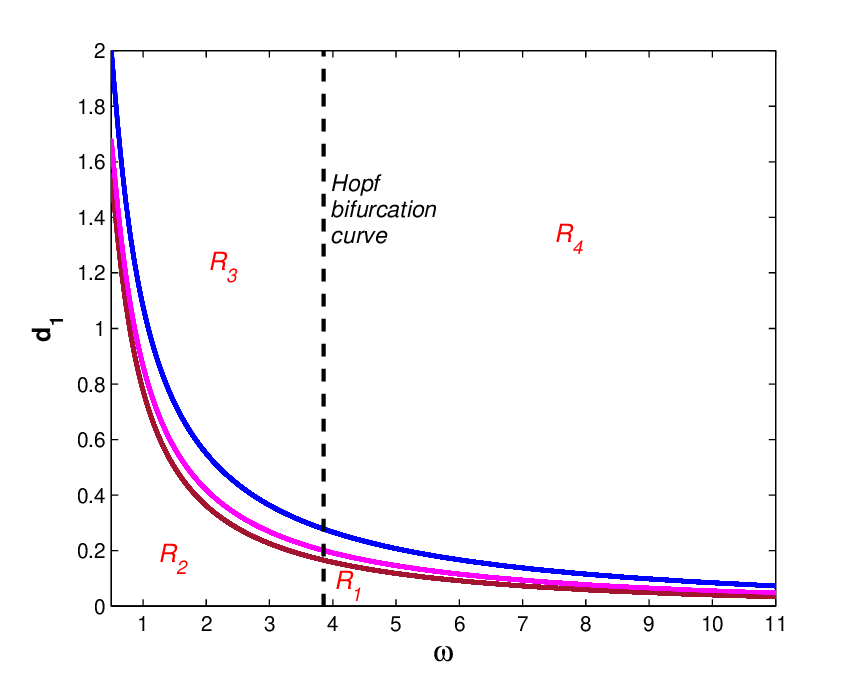}
\caption{Temporal-Hopf curve and Turing curves for local and nonlocal models. The temporal-Hopf curve is represented by black color. The maroon color curve denotes the Turing curve for the local model, and the magenta and blue color curves are the Turing curve for $\delta=5$ and $\delta=10$, respectively.} \label{fig:12}
\end{figure}

Now, we study the behaviour of the nonlocal model (\ref{eq:loc1}) for different values of the range of nonlocal interactions $\delta$. The temporal Hopf bifurcation curve is independent of $\delta$. However, the Turing bifurcation curve depends on $\delta$. We have plotted the Turing bifurcation curves in Fig. \ref{fig:12} for different values of the range of nonlocal interaction. This figure shows that the Turing curve shifts upwards with an increase in the interaction range $\delta$. This figure shows the difference between the Turing curves for local and nonlocal models to emphasize the significance of nonlocal terms in the system. Furthermore, it indicates that including nonlocal interaction expands the Turing domain by increasing the chance of occurrence of stationary patterns in the population.

Now, we find the solution characteristics for the nonlocal model in each domain formed due to the intersection of the temporal Hopf and Turing curves. For a certain nonlocal interaction $\delta$, first, we consider the region to the right of the temporal Hopf curve. The pure Turing domain is the region that lies below the Turing curve. On the other hand, the stable region lies above the Turing curve, where the solutions are homogeneous under the heterogeneous perturbations around the coexisting steady state. The region above the Turing curve and to the left of the temporal Hopf curve is called the Hopf domain. Finally, the region lying to the left of the temporal-Hopf curve and below the Turing curve is called the Turing-Hopf domain. 

\begin{figure}[ht!]
     \centering
     \begin{subfigure}{0.4\textwidth}
         \centering
         \includegraphics[width=7.5cm,height=5cm]{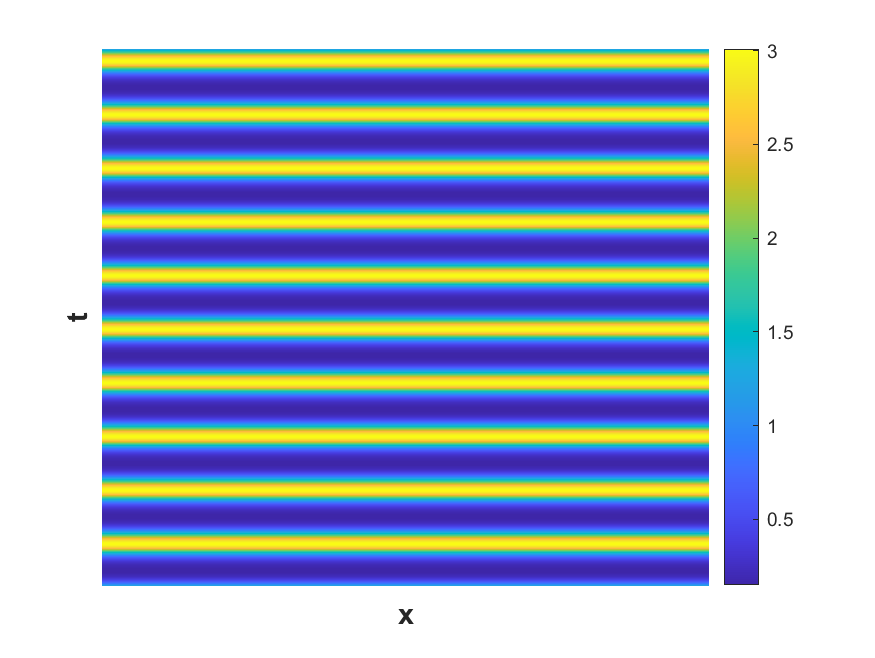}
     \end{subfigure}
     \begin{subfigure}{0.4\textwidth}
         \centering
         \includegraphics[width=7.5cm,height=5cm]{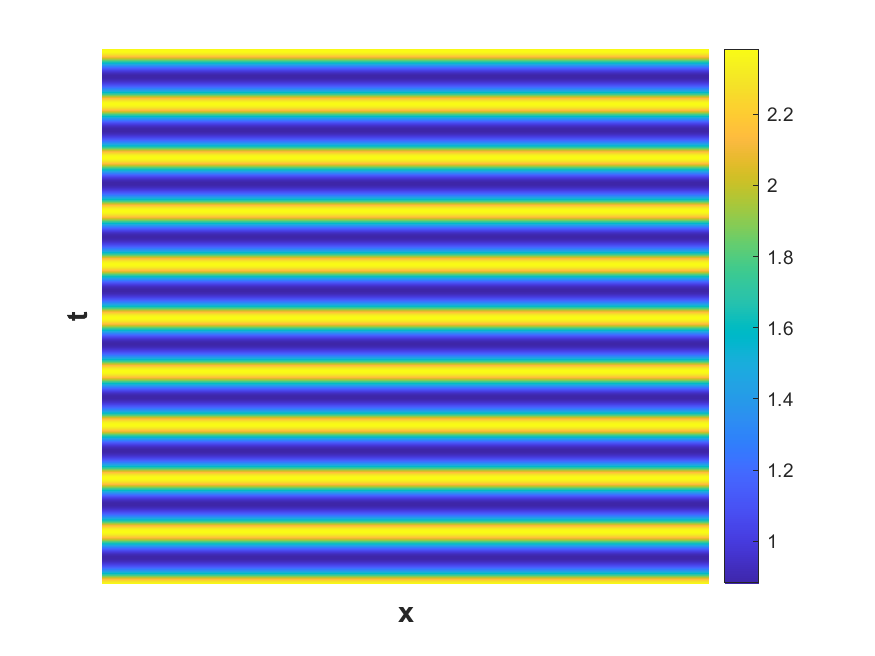}
     \end{subfigure}
 \caption{Contour plot of $u$ (left) and $v$ (right) of the nonlocal model for $\omega=2$ when $d_{1}(=2.5)$ is chosen from Hopf domain $(R_{3}$).} \label{fig:15}
\end{figure}

\begin{figure}[ht!]
     \centering
     \begin{subfigure}{0.4\textwidth}
         \centering
         \includegraphics[width=7.5cm,height=5cm]{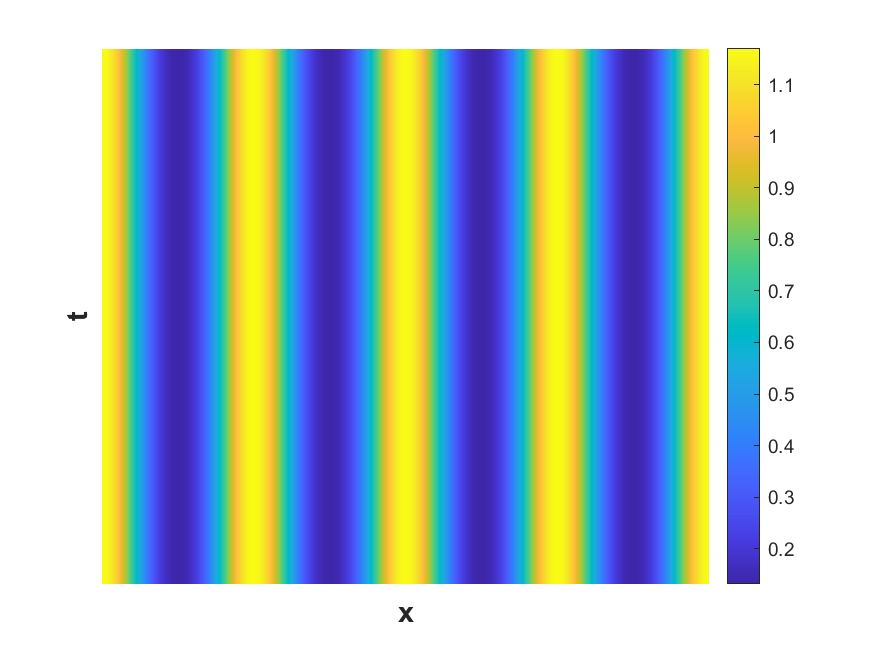}
     \end{subfigure}
     \begin{subfigure}{0.4\textwidth}
         \centering
         \includegraphics[width=7.5cm,height=5cm]{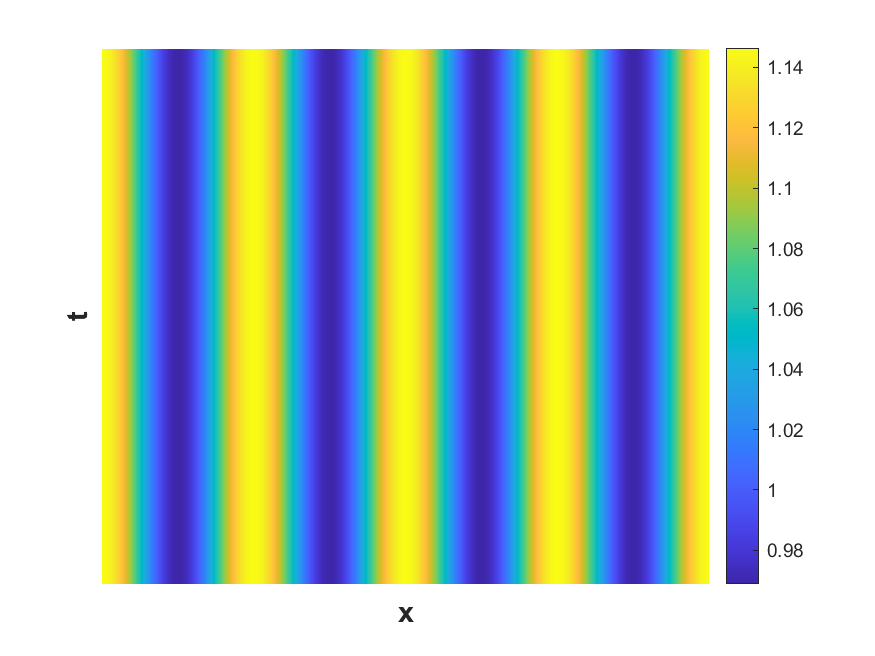}
     \end{subfigure}
\caption{Contour plot of $u$ (left) and $v$ (right) of system (\ref{eq:loc1}) for $\omega=2$ when $d_{1}(=0.38)$ is chosen from Turing-Hopf domain $(R_{2}$).}\label{fig:17}
\end{figure}

Let us first explore the solutions of the nonlocal model in the temporal Hopf unstable domain. We have chosen $\omega=2$ and $\delta=10$ and plot the corresponding solutions of the nonlocal model for $d_{1}=2.5$ and $d_{1}=0.38$ in Figs. \ref{fig:15} and \ref{fig:17}, respectively. The value $d_{1}=2.5$ lies above the Turing curve, and the other values $d_{1}=0.45$ and $d_{1}=0.38$ lie in the Turing-Hopf domain away and near the Turing curve. We mainly intend to observe if the dynamical nature changes in the presence of nonlocal interaction. The oscillatory solution occurs for system (\ref{eq:loc1}) in the Hopf domain, which is homogeneous in space but periodic in time [see Fig. \ref{fig:15} for $\omega = 2$ and $d_{1} = 2.5$]. In the Turing-Hopf domain, the dominance of Hopf mode can be seen for the nonlocal model when the parameter values lie close to the Turing curve. Moreover, a decrease in the value of $d_{1}$ ultimately gives non-homogeneous stationary patterns for the nonlocal model, which is reflected in Figure \ref{fig:17}.

\begin{figure}[ht!]
     \centering
     \begin{subfigure}{0.4\textwidth}
         \centering
         \includegraphics[width=7.5cm,height=5cm]{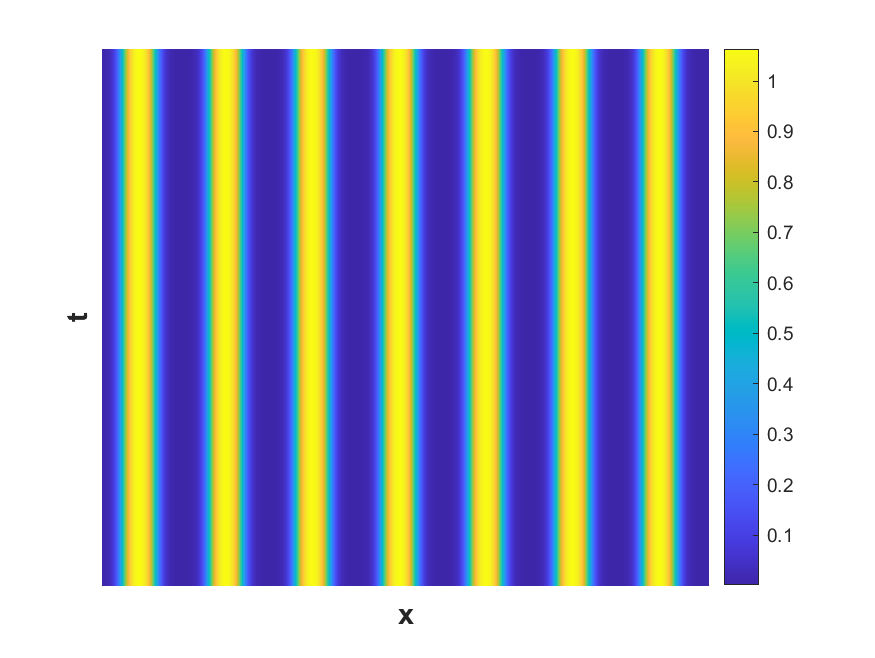}
     \end{subfigure}
     \begin{subfigure}{0.4\textwidth}
         \centering
         \includegraphics[width=7.5cm,height=5cm]{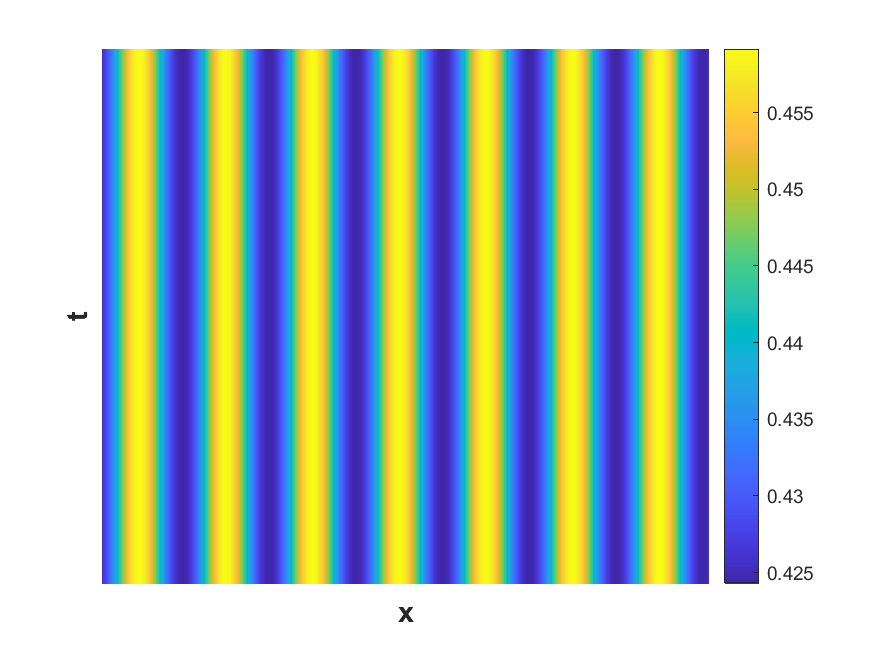}
     \end{subfigure}
\caption{Contour plot of $u$ (left) and $v$ (right) of system (\ref{eq:loc1}) for $\omega=10$ when $d_{1}(=0.01)$ is chosen from Turing domain $(R_{1}$).} \label{fig:19}
\end{figure}

For the nonlocal model, it is observed that if $d_{1}<d_{1[c]}^{T}$ holds in the right of the temporal Hopf curve, then it produces stationary Turing patterns. For instance, a Turing pattern is shown in Fig. \ref{fig:19} for $d_{1}=0.01$ and $\omega=10$. In addition, we have compared the behaviours of local and nonlocal models for the parameters that lie in the temporal Hopf unstable domain. We choose the parameter values as $\omega = 2$ and $d_{1} = 0.38$ for this. In this case, the local model shows periodic dynamics in time and homogeneous in space [see Fig. \ref{fig:8}], but the nonlocal model shows the non-homogeneous stationary solution for $\delta = 10$ [see Fig. \ref{fig:17}]; these two types of dynamics happen due to the shifting of the Turing curve. The parameter $d_{1}$ lies in the Hopf unstable domain for the local model; however, it lies in the Turing-Hopf domain for the nonlocal model with $\delta = 10$, and the Turing structure dominates the oscillation behaviour in this case. This further shows that the nonlocal interaction enhances the region's ability to get non-homogeneous stationary patterns, which is beneficial for their survival in the long run. \par

In this work, we have presented a predator-prey model incorporating the Beddington-DeAngelis functional response to account for predator interference, prey refugia, and additional food for predators during prey scarcity. The dynamics reveal that the level of prey fear acts as a stabilizing and destabilizing factor. However, coexistence is achieved when sufficient prey seek refuge, while oscillatory behaviour emerges below a critical threshold through Hopf bifurcation. The availability of additional food expands the coexistence domain, but increasing its level can lead to transcritical bifurcations, switching the stability of prey-free and interior states. Spatio-temporal analyses uncover complex oscillatory patterns and wave formations, with prey fear reducing the Turing domain, while nonlocal interactions significantly expand it, fostering spatial heterogeneity and long-range correlations. These findings underscore the intricate interplay of ecological, behavioural, and spatial mechanisms in shaping predator-prey dynamics.

\section{Conclusions} \label{sec:6}

In an ecological system, predator-prey interaction is a biological phenomenon that balances the food web. The sustainability of predatory species depends on their consumption process and search strategy for prey. This consumption process by predators depends on resource population size, availability, and the interference of other predators. Sometimes, the growth of prey becomes affected by the frequent attacks of their predator. Some works have already been published where the involvement of fear of predation is analyzed in predator-prey interactions, but nonlocal interaction is not considered there. In the current work, we have explored how incorporating nonlocal terms influences population colonization. \par
 
In this work, we have proposed a predator-prey interaction that includes psychological stress in the prey species induced by the fear of their predators. The main intention here is to elucidate the importance of this factor in the dynamic behaviour of the model, as we have assumed that the growth rate of prey is reduced due to fear of predation. Along with the fear term, we have considered the interference of predators while searching or handling their prey by choosing the Beddington-DeAngelis functional response. And, as we have assumed that the predator is provided with alternative food sources, it is evident that the prey will get the scope to move towards a predator-prohibited zone by creating prey refuge, which leaves only a fraction of them for the predators for consumption. Here we have explored how all these factors impact the dynamics of these species' interactions. It is observed in the numerical simulation that the consumption rate of predators plays an important role as the system can move to a steady prey-free state or even predator-free state from a stable coexistence state while regulating this parameter [see Fig. \ref{fig:3a}]. In fact, the detection of additional food $(\eta)$ also has the ability to regulate the dynamics as we have found stable interior equilibrium $(E^*)$ from predator-free state $(E_2)$ by decreasing its value. As the prey species creates some refugia while the predator engages with additional food, the prey species can save themselves from going extinct, and steady coexistence occurs among populations [see Fig. \ref{fig:3b}]. The impact of fear is portrayed in Fig. \ref{fig:3d} as the dual role in stabilizing and destabilizing the system. The model supports the fact that a certain range of fear levels can disturb the coexistence state of species in the environment by introducing oscillation in the system. The prey refuge $(m)$ also has the regulation ability as Fig. \ref{fig:4} shows that when the refuge parameter lies below a threshold value, the population oscillates, but we get coexistence while $m$ exceeds the Hopf threshold. Not only that, but the presence of additional food increases the chances of such coexistence. Now, the fear level has been found to be a stabilizing as well as destabilizing factor in this model [see Fig. \ref{fig:3d}]. Though a stable coexisting state found for a very small as well as large value of fear $(\omega<\omega_{[H_1]}\ \mbox{and}\ \omega>\omega_{[H_2]})$, but oscillation is observed when it lies within a range $\omega\in(\omega_{[H_1]},\ \omega_{[H_2]})$. It indicates that a certain amount of fear is needed in the system for the population to coexist.  \par

In the later part of the work, it is assumed that the species can move in one direction described by the spatio-temporal model. The analysis reveals that the diffusion coefficient for the prey species starts to decrease with increasing fear level $(\omega)$ [see Fig. \ref{fig:5}]. It indicates that the prey, out of fear of being hunted, will avoid moving in the mentioned direction. In fact, the increase in $\omega$ shrinks the region of the Turing domain, reducing the chances of non-homogeneous pattern formation. As the species are not always homogeneously distributed over a domain, this shrinkage may not be favourable for persistence. Moreover, Fig. \ref{fig:sp2} shows that an increase in either prey refuge or additional food reduces the spatial extent of species colonization by narrowing the Turing domain. This implies that higher values of these parameters lower the prey diffusion threshold $(d_{1[c]})$ needed for the emergence of Turing patterns, thereby limiting the parameter space conducive to patch formation. From an ecological perspective, an increase in prey refuge and additional food reduces environmental heterogeneity, potentially leading to more uniform species distributions. This may enhance prey survival but could also diminish spatial complexity, which is often crucial for maintaining biodiversity and ecosystem resilience. Furthermore, incorporating nonlocal terms in the system shifts the Turing curve upwards, expanding the Turing domain. This means that the species can be colonized with an increasing range of nonlocal interactions, which will benefit both species' survival in the future. \par

Refining the proposed model requires addressing both immediate enhancements and more complex modifications that necessitate further investigation. While certain aspects can be incorporated in the short term, others require a deeper theoretical and empirical foundation to ensure ecological validity. In the short term, the proposed model can be refined by incorporating group defense mechanisms in prey growth, thereby enhancing its ecological realism. Additionally, the assumption of a fixed proportion of prey successfully seeking refuge can be replaced with a predator-dependent refuge function, allowing for a more dynamic representation of prey behaviour. Another immediate improvement involves introducing environmental stochasticity through white Gaussian noise, which would account for random fluctuations in ecological interactions. In the long term, more complex modifications require further theoretical and empirical investigation. One such refinement is the incorporation of the carryover effect, wherein the influence of fear on prey behaviour extends across generations, necessitating a more detailed analysis of inter-generational dynamics. Furthermore, since the impact of predation risk on prey behaviour develops over time, introducing a time-delay component would provide a more realistic depiction of this process. However, this modification requires a rigorous mathematical framework to assess its effects on system stability and dynamics. Addressing both short-term and long-term considerations will contribute to a more comprehensive and ecologically relevant model of predator-prey interactions.


\subsection{Acknowledgements}
The authors are grateful to the NSERC and the CRC Program for their support. RM is also acknowledging the support of the BERC 2022–2025 program and the Spanish Ministry of Science, Innovation and Universities through the Agencia Estatal de Investigacion (AEI) BCAM Severo Ochoa excellence accreditation SEV-2017–0718. This research was enabled in part by support provided by SHARCNET (\url{www.sharcnet.ca}) and Digital Research Alliance of Canada (\url{www.alliancecan.ca}).

\subsection{Data Availability Statement}
The data used to support the findings of the study are available within the article.

\subsection{Conflict of Interest}
This work does not have any conflict of interest.

\begin{appendices}

\section{}\label{appendixA}

\subsection{Proof of positivity and boundedness of variables of model (\ref{eq:det1})}\label{appendixA1}
Functions on the right-hand side of the system (\ref{eq:det1}) are continuous and locally Lipschitzian (as they are polynomials and rationals in $(u,v)$), so there exists a unique solution $(u(t),v(t))$ of the system with positive initial conditions $(u(0), v(0))>0$ on $[0,\tau],$ where $0<\tau<+\infty$ \cite{hale1977}. From the first, and second equation of (\ref{eq:det1}) we have
\begin{align*}
\frac{du}{dt}&=u\left[\frac{r}{1+\omega v}-d-pu-\frac{a(1-m)v}{b+\alpha\eta A+(1-m)u+lv}\right]=u\psi_{1}(u,v) \\
\Rightarrow u(t)&=u(0)\exp\left[\int^t_0 \psi_{1}(u(z),v(z))\,dz\right]> 0, \ \textrm{for} \ u(0)> 0.
\end{align*}
\begin{align*}
\textrm{And,}\ \ \frac{dv}{dt}&=v\left[\frac{ca\{(1-m)u+\eta A\}}{b+\alpha\eta A+(1-m)u+lv}-\mu_{1}-\mu_{2}v\right]=v\psi_{2}(u,v) \\
\Rightarrow v(t)&=v(0)\exp\left[\int^t_0 \psi_{2}(u(z),v(z))\,dz\right]> 0, \ \textrm{for} \ v(0)> 0.
\end{align*}
So, the solutions of the system (\ref{eq:det1}) are feasible with time. Now, the first equation of (\ref{eq:det1}) gives:
\begin{equation*}
\begin{aligned}
\frac{du}{dt}&=\frac{ru}{1+\omega v}-du-pu^{2}-\frac{a(1-m)uv}{b+\alpha\eta A+(1-m)u+lv} \leq (r-d)u\left[1-\frac{pu}{r-d}\right] \\
\displaystyle \Rightarrow & \limsup_{t\rightarrow \infty}u(t)\leq \frac{r-d}{p}.
\end{aligned}
\end{equation*}
 
\noindent Consider, $W(t)= u(t)+\frac{1}{c}v(t)$. Then for any $\eta>0$ we have
\begin{equation*}
\begin{aligned}
\textrm{So,}\ \frac{dW}{dt}+\eta W&= \left(\frac{du}{dt}+\frac{1}{c}\frac{dv}{dt}\right)+\eta u+\frac{\eta}{c}v \\
& \leq u[(r-d)-pu+\eta]+\frac{a\eta Av}{b+\alpha\eta A+(1-m)u+lv}-\frac{\mu_{1}}{c}v+\frac{\eta}{c}v \\
& \leq \frac{(r-d+\eta)^{2}}{4p}+\frac{a\eta A}{l}-\frac{v}{c}(\mu_{1}-\eta).
\end{aligned}
 \end{equation*}
Choosing sufficiently small $\eta\ll\mu_{1}$, we obtain
 \begin{equation*}
 \frac{dW}{dt}+\eta W \leq \left[\frac{(r-d+\eta)^{2}}{4p}+\frac{a\eta A}{l}\right]=M.    
 \end{equation*}
By applying Gronwall’s inequality, we have
then $\displaystyle 0\leq W(t)\leq \frac{M}{\eta}(1-\exp(-\eta t))+ W(u(0),v(0))\exp(-\eta t)$. \\
As $\displaystyle t\rightarrow\infty,\ 0<W(t)\leq M/\eta+\epsilon$ for sufficiently small $\epsilon>0$. So, the solutions of system (\ref{eq:det1}) enter into the bounded region: 
\[\Omega = \left\{(u,v):\ 0\leq u(t)\leq (r-d)/p;\ 0\leq W(t)\leq  M/\eta+\epsilon,\ \epsilon>0\right\}.\]

\subsection{Proofs of local stability conditions of the equilibrium points of system (\ref{eq:det1})}\label{appendixA2}

The Jacobian matrix corresponds to $E_{0}(0,0)$ is given as 
\begin{equation*}
\textbf{J}|_{E_{0}}=\left(
\begin{array}{cc}
r-d & 0 \\
0 & \frac{ca\eta A}{s}-\mu_{1}
\end{array}
\right),
\end{equation*}
which gives $\lambda_{1}=r-d$ and $\lambda_{2}=ca\eta A/s-\mu_{1}$. It means $\lambda_{2}<0$ when $ca\eta A<\mu_{1}s$ but $\lambda_{1}>0$ always (as the system is bounded). As one of the eigenvalues is positive, and hence, $E_{0}$ is an unstable equilibrium point.

\noindent Moreover, the Jacobian matrix at $E_{1}(\overline{u},0)$ is
\begin{equation*}
\textbf{J}|_{E_{1}}=\left(
\begin{array}{cc}
b_{11} & b_{12} \\
0 & b_{22}
\end{array}
\right)=\begin{pmatrix}
   -p\overline{u} & -r\omega\overline{u}-\frac{a(1-m)\overline{u}}{s+(1-m)\overline{u}} \\
0 & \frac{ca\{(1-m)\overline{u}+\eta A\}}{s+(1-m)\overline{u}}-\mu_{1} 
\end{pmatrix}.
\end{equation*}

The eigenvalues are the roots of the equation: $\lambda^{2}+C_{1}\lambda+C_{2}=0$,
where $C_{1}=-(b_{11}+b_{22})$ and $C_{2}=b_{11}b_{22}$. So, the equation has roots with negative real parts if $C_{1}>0$ and $C_{2}>0$, and this occurs when $b_{22}<0$, i.e., $\displaystyle (ca-\mu_{1})(1-m)\overline{u}<\mu_{1}s-ca\eta A$. 

Again, the Jacobian matrix corresponds to $E_{2}(0,\widetilde{v})$ is denoted as
\begin{equation*}
\textbf{J}|_{E_{2}}=\left(
\begin{array}{cc}
b_{11} & 0 \\
b_{21} & b_{22}
\end{array}
\right)=\begin{pmatrix}
\frac{r}{1+\omega\widetilde{v}}-\frac{a(1-m)\widetilde{v}}{s+l\widetilde{v}}-d & 0 \\
\frac{ca(1-m)\widetilde{v}\{b+l\widetilde{v}+\eta A(\alpha-1)\}}{(s+l\widetilde{v})^{2}} & -\mu_{2}\widetilde{v}-\frac{ca\eta Al\widetilde{v}}{(s+l\widetilde{v})^{2}}
\end{pmatrix}.
\end{equation*}

The eigenvalues are the roots of the equation: $\lambda^{2}+D_{1}\lambda+D_{2}=0,$
where $D_{1}=-(b_{11}+b_{22})$ and $D_{2}=b_{11}b_{22}$. The equation has roots with negative real parts if $D_{1}>0$ and $D_{2}>0$ and this occurs when $b_{11}<0$, i.e., $\omega\{dl+a(1-m)\}\widetilde{v}^{2}+\{ds\omega+a(1-m)-(r-d)l\}\widetilde{v}-(r-d)s>0$. 

Lastly, the Jacobian matrix corresponds to $E^{*}(u^{*}, v^{*})$ is given as
\begin{equation*}
\textbf{J}(E^*)=\textbf{J}|_{E^{*}}=\left(
\begin{array}{cc}
a_{11} & a_{12} \\
a_{21} & a_{22}
\end{array}
\right),
\end{equation*}
where $\displaystyle a_{11}=-pu^{*}+\frac{a(1-m)^{2}u^{*}v^{*}}{[s+(1-m)u^{*}+lv^{*}]^{2}},\ a_{12}=-r\omega u^{*}-\frac{a(1-m)u^{*}[s+(1-m)u^{*}]}{[s+(1-m)u^{*}+lv^{*}]^{2}},\\ a_{21}=\frac{ca(1-m)v^{*}[b+lv^{*}+\eta A(\alpha-1)]}{[s+(1-m)u^{*}+lv^{*}]^{2}}$ and $\displaystyle a_{22}=-\mu_{2}v^{*}-\frac{ca\{(1-m)u^{*}+\eta A\}lv^{*}}{[s+(1-m)u^{*}+lv^{*}]^{2}}$. \\
The characteristic equation corresponding to $\textbf{J}|_{E^{*}}$ is given as follows: 
\mathcenter
\begin{equation}\label{eq:2.4}
\lambda^{2}+F_{1}\lambda+F_{2}=0,
\end{equation}
where $F_{1}=-\mbox{tr}(\textbf{J}(E^{*}))=-(a_{11}+a_{22})$ and $F_{2}=\det(\textbf{J}(E^{*}))=a_{11}a_{22}-a_{12}a_{21}$. So, by Routh- Hurwitz criteria, the equilibrium point will be locally asymptotically stable if the equation has roots with negative real parts, i.e., $F_{1}>0$ and $F_{2}>0$, i.e., 
\begin{equation} \label{eq:2.4_1}
    a_{11}+a_{22}<0\ \textrm{and}\ a_{11}a_{22}>a_{12}a_{21}.
\end{equation}
So, the local asymptotic stability of $E^{*}$ depends on the conditions: $av^{*}[(1-m)^{2}u^{*}-lc(1-m)u^{*}-lc\eta A]<(pu^{*}+\mu_{2}v^{*})K^{2},$ and $p\mu_{2}K^{4}+[ca\{pl\{(1-m)u^{*}+\eta A\}+r\omega(1-m)\{b+lv^{*}+\eta A(\alpha-1)\}\}-a\mu_{2}(1-m)^{2}v^{*}]K^{2}+ca^{2}(1-m)^{2}[\{s+(1-m)u^{*}\}\{b+lv^{*}+\eta A(\alpha-1)\}+(s-\eta A)lv^{*}]>0$, where $K=[s+(1-m)u^{*}+lv^{*}]>0$.

\subsection{Proofs of local bifurcations around the equilibrium points of system (\ref{eq:det1})}\label{appendixA3}

Let us consider $\textbf{V}= (v_{1},v_{2})^{T}$ and $\textbf{W}= (w_{1},w_{2})^{T}$, respectively be the eigenvectors of $\textbf{J}|_{\left(eq.\ point\right)}$ and $\textbf{J}|^{T}_{\left(eq.\ point\right)}$ for a zero eigenvalue at the equilibrium point. Let $\overline{\textbf{F}}=(\overline{F}_{1},\overline{F}_{2})^{T},$ where 
\mathleft
\begin{align*}
 \overline{F}_{1}&=\frac{ru}{1+\omega v}-du-pu^{2}-\frac{a(1-m)uv}{b+\alpha\eta A+(1-m)u+lv},\\
 \overline{F}_{2}&=\frac{ca\{(1-m)u+\eta A\}v}{b+\alpha\eta A+(1-m)u+lv}-\mu_{1}v-\mu_{2}v^{2}.
\end{align*}

Now, the Jacobian matrix at $E_{1}$ is given by
\mathcenter
\begin{equation*}
\textbf{J}|_{E_{1}}=\begin{pmatrix}
b_{11} & b_{12}\\
0 & b_{22}
\end{pmatrix}=\begin{pmatrix}
   -p\overline{u} & -r\omega\overline{u}-\frac{a(1-m)\overline{u}}{s+(1-m)\overline{u}} \\
0 & \frac{ca\{(1-m)\overline{u}+\eta A\}}{s+(1-m)\overline{u}}-\mu_{1} 
\end{pmatrix}.
\end{equation*}

The eigenvalues are the roots of the equation $\lambda^{2}-(b_{11}+b_{22})\lambda+b_{11}b_{22}=0$, which gives the roots with negative real parts. Let $a_{[TC_1]}$ be the value of $a$ such that $(ca-\mu_{1})(1-m)\overline{u}=\mu_{1}s-ca\eta A$ so that $\textbf{J}|_{E_{1}}$ has a simple zero eigenvalue at $a_{[TC_1]}$. So, at $a=a_{[TC_1]}:$
\begin{equation*}
\textbf{J}|_{E_{1}}=\begin{pmatrix}
b_{11} & b_{12} \\
0 & 0
\end{pmatrix}.
\end{equation*}
The calculations give $\textbf{V}=(v_{1},v_{2})^{T}$ where $v_{1}=-b_{12},\ v_{2}=b_{11}$ and $\textbf{W}=(0,1)^{T}.$ Therefore,
\begin{equation*}
\begin{aligned}
\Omega_{1}&= \textbf{W}^{T}.\overline{\textbf{F}}_{a}(E_{1}, a_{[TC_1]})=\frac{c\{(1-m)\overline{u}+\eta A\}v}{\{s+(1-m)u+lv\}}\bigg|_{E_{1}}=0, \\
\Omega_{2}&= \textbf{W}^{T}\left[D\overline{\textbf{F}}_{a}(E_{1}, a_{[TC_1]})\textbf{V}\right]=\frac{c\{(1-m)\overline{u}+\eta A\}b_{11}}{\{s+(1-m)\overline{u}\}}\bigg|_{E_{1}}\neq 0, \\
\textrm{and}\ \Omega_{3}&= \textbf{W}^{T}\left[D^{2}\overline{\textbf{F}}(E_{1},a_{[TC_1]})(\textbf{V},\textbf{V})\right]\\
&=\frac{-2ca(1-m)\{b+\eta A(\alpha-1)\}b_{11}b_{22}}{\{s+(1-m)\overline{u}\}^{2}}-2\left\{\mu_{2}+\frac{cal\{(1-m)\overline{u}+\eta A\}}{\{s+(1-m)\overline{u}\}^{2}}\right\}b_{11}^{2} \neq0. \\
\end{aligned}
\end{equation*}
\mathcenter
Then, by Sotomayor's Theorem, the system undergoes a transcritical bifurcation around $E_{1}$ at $a=a_{[TC_1]}.$

Similarly, we can prove that the system exhibits another transcritical bifurcation around $E_{2}$ at $a=a_{[TC_2]}$, where $a_{[TC_2]}$ is the positive root of the equation $\omega\{dl+a(1-m)\}\widetilde{v}^{2}+\{ds\omega+a(1-m)-(r-d)l\}\widetilde{v}-(r-d)s=0$.  

Now, for the existence of Hopf bifurcation around $E^{*}$, at $\omega=\omega_{[H]}$, the characteristic equation of system (\ref{eq:det1}) at $E^{*}$ is $(\lambda^{2}+F_{2})=0$ and so, the equation has a pair of purely imaginary roots $\lambda_{1}=i\sqrt{F_{2}}$ and $\lambda_{2}=-i\sqrt{F_{2}}$ where $F_{2}(\omega)$ is a continuous function of $\omega$. \\
In the small neighbourhood of $\omega_{[H]},$ the roots are $\lambda_{1}=p_{1}(\omega)+ip_{2}(\omega)$ and $\lambda_{2}=p_{1}(\omega)-ip_{2}(\omega)$ (where $p_{1}$ and $p_{2}$ are $\mathcal{C}^{1}$ functions in $\mathbb{R}$). \\
To show the transversality condition, we need to check $\displaystyle \left(\frac{d}{d\omega}[Re(\lambda_{i}(\omega))]\right)\bigg|_{\omega=\omega_{[H]}}\neq 0,$ for $i=1,2.$ \\
Put $\lambda(\omega)=p_{1}(\omega)+ip_{2}(\omega)$ in (\ref{eq:2.4}), we get
\mathcenter
\begin{equation} \label{eq:2.5}
(p_{1}+ip_{2})^{2}+F_{1}(p_{1}+ip_{2})+F_{2}=0.
\end{equation}
Differentiating with respect to $\omega$, we get
\begin{equation*}
2(p_{1}+ip_{2})(\dot{p_{1}}+i\dot{{p_{2}}})+F_{1}(\dot{p_{1}}+i\dot{p_{2}})+\dot{F_{1}}(p_{1}+ip_{2})+\dot{F_{2}}=0.
\end{equation*}
Comparing the real and imaginary parts from both sides, we have \\
\begin{equation} \label{eq:2.6}
(2p_{1}+F_{1})\dot{p_{1}}-(2p_{2})\dot{p_{2}}+(\dot{F_{1}}p_{1}+\dot{F_{2}}) = 0, \\
\end{equation}
\begin{equation} \label{eq:2.7}
(2p_{2})\dot{p_{1}}+(2p_{1}+F_{1})\dot{p_{2}}+\dot{F_{1}}p_{2}=0.
\end{equation}
Solving we get, $\displaystyle \dot{p_{1}}=\frac{-2p_{2}^{2}\dot{F_{1}}-(2p_{1}+F_{1})(\dot{F_{1}}p_{1}+\dot{F_{2}})}{(2p_{1}+F_{1})^{2}+4p_{2}^{2}}$. \\
At, $p_{1}=0,\ p_{2}=\pm \sqrt{F_{2}}:\ \displaystyle \dot{p_{1}}=\frac{-2\dot{F_{1}}F_{2}-F_{1}\dot{F_{2}}}{F_{1}^{2}+4F_{2}}\neq 0$. Hence, this completes the proof.

\section{}\label{appendixB}

The homogeneous steady states for both the local model (\ref{eq:diff1}) and the nonlocal model (\ref{eq:loc1}) are the same. Let us consider $E^{*}(u^{*},v^{*})$ as the coexisting homogeneous steady-state. Now, perturbing the system around $(u^{*},v^{*})$ by 
\begin{align*}
    \begin{pmatrix}
        u \\
        v
    \end{pmatrix}=\begin{pmatrix}
        u^{*} \\
        v^{*}
    \end{pmatrix}+\epsilon \begin{pmatrix}
        u_{1} \\
        v_{1}
    \end{pmatrix}e^{\lambda t+ikx}
\end{align*}
with $|\epsilon|<<1$ and substituting these values in system (\ref{eq:loc1}) the linearization takes the form:
\begin{equation} \label{eq:4.2}
\overline{\textbf{J}}_{k}\begin{bmatrix}
u_{1}\\
v{1}
\end{bmatrix}\equiv \begin{bmatrix}
 a_{11}-d_{1}k^{2}-\lambda & A_{12} \\
 a_{21} & a_{22}-d_{2}k^{2}-\lambda 
\end{bmatrix}
\begin{bmatrix}
    u_{1} \\
    v_{1}
\end{bmatrix}
=\begin{bmatrix}
        0 \\
        0
    \end{bmatrix},
\end{equation}
where $a_{11},\ a_{12},\ a_{21}$ and $a_{22}$ are mentioned in the proof of Theorem \ref{th2.2} and
$$\displaystyle A_{12}=-\left\{u^{*}\frac{\partial f_{1}}{\partial v}\bigg|_{(u^{*},v^{*})}+\frac{r\omega u^{*}}{(1+\omega v^{*})^{2}}\frac{\sin k\delta}{k\delta}\right\}=a_{12}+\frac{r\omega u^{*}}{(1+\omega v^{*})^{2}}\left\{1-\left(\frac{\sin k\delta}{k\delta}\right)\right\}.$$
Now, we are interested in finding the non-trivial solution of the system (\ref{eq:4.2}), so $\lambda$ must be a zero of $\det(\overline{\textbf{J}}_{k})=0$, where $\overline{\textbf{J}}_{k}$ is the coefficient matrix of the system (\ref{eq:4.2}). Now 
\begin{align*}
\det(\overline{\textbf{J}}_{k})=\lambda^{2}-\Gamma(k,d_{1},\delta)\lambda+\Delta(k,d_{1},\delta),
\end{align*}
where $\Gamma(k,d_{1},\delta)=(a_{11}+a_{22})-k^{2}(d_{1}+d_{2})$ and $\Delta(k,d_{1},\delta)=d_{1}d_{2}k^{4}-(a_{11}d_{2}+a_{22}d_{1})k^{2}+(a_{11}a_{22}-a_{21}A_{12})$. So, $\det(\overline{\textbf{J}}_{k})=0$ will give
\begin{align*}
\lambda(k^{2})=\frac{\Gamma(k,d_{1},\delta)\pm\sqrt{(\Gamma(k,d_{1},\delta))^{2}-4\Delta(k,d_{1},\delta)}}{2}.
\end{align*}
The homogeneous steady-state $E^{*}(u^{*},v^{*})$ is stable if $\Gamma(k,d_{1},\delta)<0$ and $\Delta(k,d_{1},\delta)>0$ holds for all $k$ for some fixed $d_{1}$ and $\delta$. Now, if the local model is stable, then we already have $\Gamma(k,d_{1},\delta)<0$ here. So, the instability of the coexisting homogeneous steady-state depends on the sign of $\Delta(k,d_{1},\delta)$ only. Moreover, Turing instability occurs if $\Gamma(k,d_{1},\delta)<0$ holds for all $k$ and $\Delta(k,d_{1},\delta)=0$ holds for a unique $k$. 
$\Gamma(k,d_{1},\delta)$ and $\Delta(k,d_{1},\delta)$ depending on the parameter $\delta$, and hence it plays an important role for the above instabilities. Therefore, we find these instability conditions by fixing $\delta$. We assume that the equilibrium point $E^{*}$ is locally asymptotically stable for the temporal model.

To find the condition of Turing instability of the nonlocal model, we need to find a value of $d_{1}$ such that $\Delta(k,d_{1},\delta)=0$ for a fixed $k$ and $\Gamma(k,d_{1},\delta)<0$ for all k. Also, for all $d_{1}$, we have got $\Delta(k,d_{1},\delta)>0$ when $k$ is sufficiently small as well as a large quantity. So, $\Delta(k,d_{1},\delta)=0$ holds for a unique $k$ when 
$$\Delta(k,d_{1},\delta)=0\ \ \textrm{and}\ \ \frac{\partial \Delta(k,d_{1},\delta)}{\partial k}=0$$ hold, i.e., 
\begin{subequations}\label{eq:4.3}
\begin{align} 
d_{1}&=\frac{d_{2}a_{11}k^{2}-(a_{11}a_{22}-a_{12}a_{21})+\frac{a_{21}r\omega u^{*}}{(1+\omega v^{*})^{2}}\left(1-\frac{\sin k\delta}{k\delta}\right)}{k^{2}(d_{2}k^{2}-a_{22})}, \label{eq:4.3a} \\
\mbox{and}\ 4d_{1}d_{2}k^{3}&-2(d_{2}a_{11}+d_{1}a_{22})k+\frac{a_{21}r\omega u^{*}}{k\delta(1+\omega v^{*})^{2}}\left(\delta \cos k\delta-\frac{\delta \sin k\delta}{k\delta}\right)=0. \label{eq:4.3b}
\end{align}
\end{subequations}
From (\ref{eq:4.3}), eliminating $d_{1}$ leads to the following transcendental equation
\begin{equation} \label{eq:4.4}
\begin{aligned}
2a_{11}d_{2}^{2}k^{4}-2(2d_{2}k^{2}-a_{22})\left[(a_{11}a_{22}-a_{12}a_{21})-\frac{a_{21}r\omega u^{*}}{(1+\omega v^{*})^{2}}\left\{1-\left(\frac{\sin k\delta}{k\delta}\right)\right\}\right]+\\
\frac{a_{21}r\omega u^{*}(d_{2}k^{2}-a_{22})}{(1+\omega v^{*})^{2}}\left(\cos k\delta-\frac{\sin k\delta}{k\delta}\right)=0,
\end{aligned}
\end{equation}
which needs to be solved numerically for a fixed value of $\delta$ to obtain the critical wave number $k_{\min}^{T}$. Here, we may get multiple solutions of (\ref{eq:4.4}) for a large value of $\delta$. Out of these multiple values of $k$, we choose $k_{\min}^{T}$ for which $\Delta(k,d_{1},\delta)=0$ holds for a unique $k$ [see Fig. \ref{fig:6}(a)]. Substitution of this value in (\ref{eq:4.3a}) will give the critical diffusion coefficient $d_{1[c]}^{T}$. Here $d_{1}<d_{1[c]}^{T}$ leads to $\Delta(k,d_{1},\delta)<0$, so, Turing instability occurs for $d_{1}<d_{1[c]}^{T}$.

\end{appendices}

\bibliography{P_References}

\end{document}